\documentclass[twocolumn,apj,numberedappendix,twocolappendix]{openjournal}
\pdfoutput=1

\usepackage{booktabs}
\usepackage{capt-of}
\usepackage{amsmath}
\usepackage{algpseudocode}
\usepackage[colorlinks=true,linkcolor=blue,citecolor=blue,urlcolor=blue]{hyperref}
\usepackage{orcidlink}

\newcommand{\paynezero}{\textsc{Payne Zero}}
\newcommand{\tcell}[2]{\parbox[t]{#1}{\raggedright #2}}
\newcommand{\compacttablefont}{\footnotesize}

\shorttitle{Stellar Spectra from Physical Models in Seconds}
\shortauthors{Ting \& Kim}

\makeatletter
\def\frontmatter@title@below{\vspace*{-2.63\baselineskip}\vspace*{0.37in}}
\def\frontmatter@above@affilgroup{\vspace*{0.18in}}
\journalinfo{Accepted for publication in the Open Journal of Astrophysics}
\submitted{Version \today}
\makeatother

\begin{document}

\title{The Payne Zero Project I\\Stellar Spectra from Physical Models in Seconds}

\author{Yuan-Sen Ting\,\orcidlink{0000-0001-5082-9536}}
\affiliation{Department of Astronomy, The Ohio State University, 140 West 18th Avenue, Columbus, OH 43210, USA}
\affiliation{Center for Cosmology and AstroParticle Physics (CCAPP), The Ohio State University, Columbus, OH 43210, USA}
\affiliation{Max-Planck-Institut f\"ur Astronomie, K\"onigstuhl 17, D-69117 Heidelberg, Germany}

\author{Elliot M. Kim}
\affiliation{Department of Computer Science, Cornell University, Ithaca, NY 14853, USA}

\begin{abstract}
Modern stellar surveys measure millions of spectra, yet one self-consistent atmosphere and spectrum can require tens of minutes. This cost has motivated grids, spectral emulators, and data-driven models. We present \paynezero, which reorganizes one-dimensional LTE Kurucz calculations for GPU-native synthesis and multicore atmosphere iteration, and validate it against the original Fortran programs. A 300--1000~nm solar spectrum sampled at $R_{\rm grid}=300{,}000$ takes about 14~s on an NVIDIA H100 GPU, while the APOGEE 1500--1700~nm interval takes about 1~s. Physical atmosphere iterations take 2--5~s on 16 AMD CPU threads, and learned initializers reduce the iterations required for convergence. Final spectra remain in practical parity across the tested dwarf and giant regimes. These speeds place direct synthesis inside an optimizer without a label-to-flux spectral emulator. We demonstrate direct many-element fitting of reduced APOGEE spectra and recover multi-element abundance trends broadly consistent with the survey catalog. GPU-resident velocity shifts, broadening, line-spread-function convolution, and detector sampling add negligible cost relative to synthesis. The direct-synthesis search takes less than one minute per star on an H100, while atmosphere verification runs independently on multicore CPUs. The same computational graph calibrates more than $10^5$ oscillator-strength and damping corrections jointly against the Sun and Arcturus in about one minute on an H100. \paynezero\ therefore brings direct physical fitting and atomic-data calibration to survey scale. The code is available at \url{https://github.com/tingyuansen/payne-zero}.
\end{abstract}


\section{Introduction}\label{sec:introduction}

Large spectroscopic surveys have measured, are measuring, or will soon measure spectra for hundreds of thousands to millions of stars. Major Milky Way programs include APOGEE, GALAH, Gaia--ESO, LAMOST, Gaia RVS, SDSS-V, DESI, WEAVE, 4MOST, and PFS \citep{Cui2012,Dalton2012,Gilmore2012,Takada2014,DeSilva2015, Kollmeier2017,Majewski2017,Chiappini2019,Cooper2023,Creevey2023}. Their scientific return depends on turning each spectrum into stellar properties, including effective temperatures, surface gravities, chemical abundances, and radial velocities. The speed and accuracy of the forward model set a basic limit on that analysis.

A physical forward model combines three stages. A model atmosphere describes the temperature, pressure, electron density, and chemical populations with depth. Spectral synthesis combines this structure with a line list to calculate the emergent radiation. An instrument model maps the intrinsic spectrum to the observed pixels. The atmosphere supplies the thermodynamic and chemical state used by synthesis, so the same composition controls both the atmospheric structure and the emergent spectrum.

The requested abundance pattern enters the equation of state and opacity as a whole. Magnesium, for example, changes both its own lines and the electron supply, while electron donors alter the H$^-$ continuum and thereby features from other species \citep{John1988,Meszaros2012,Matsuno2024}. Carbon, nitrogen, and oxygen redistribute atoms among molecules such as CO, CN, and OH and can therefore change spectral features not belonging to the perturbed element \citep{Tsuji1973,Ting2018Oxygen}. Line blanketing from many metals changes the radiation field and therefore the temperature structure \citep{Gustafsson2008}. The opacity-sampling atmosphere solver recomputes the populations and opacity for the requested mixture \citep{Kurucz1996}.

Runtime has long determined which parts of this calculation enter an analysis. Classical equivalent-width work can obtain a model atmosphere by interpolation within a precomputed grid and synthesize a small region around an isolated line \citep{CastelliKurucz2003}. Long-standing and modern codes and analysis frameworks include MOOG, Turbospectrum, FERRE, iSpec, Korg, SME, and PySME \citep{ValentiPiskunov1996,Plez2012,MOOG2012,BlancoCuaresma2014, PiskunovValenti2017,FERRE2023,Wehrhahn2023,Wheeler2023}. Full-spectrum survey analyses instead draw information from broad wavelength intervals containing isolated lines, blends, and, in cool stars, molecular bands \citep{ValentiFischer2005,Ting2017Prospects}.

For each new stellar state, the original Kurucz workflow stages atmosphere and synthesis through serial programs and intermediate files \citep{Kurucz2005}. The validated Python reimplementation of \citet{KimTing2026} provides the direct software and physics lineage for our redesign. In the benchmarks below, the label-dependent atmosphere setup, physical iterations, and broad synthesis in the original Fortran workflow together take tens of minutes. Repeating that calculation inside a full-spectrum optimizer can then take hours for one star.

Spectral emulators avoid this cost by interpolating synthetic libraries, a practice whose atmosphere- and flux-interpolation errors can be measured directly \citep{MeszarosAllendePrieto2013}. Their implementations include probabilistic interpolation, as in Starfish, and neural interpolation, as in The Payne \citep{Czekala2015,Ting2019}. An emulator still has a finite training range and interpolation error, and a new physical model or label set requires a new grid \citep{Rozanski2025a}. The training problem becomes harder when many individual abundances vary because the volume of label space grows rapidly \citep{Casey2016,Ting2016}.

Data-driven models such as The Cannon and its regularized and cross-survey extensions, StarNet, and astroNN instead learn label--flux or label-inference relations from observed reference spectra \citep{Ness2015,Casey2016,Ho2017,Fabbro2018,LeungBovy2019}. Correlated training labels can mix physical sensitivity with astrophysical correlations \citep{Ting2017LAMOST,Xiang2019}. Related domain-adaptation methods such as Cycle-StarNet map synthetic and observed spectral domains to reduce repeatable discrepancies \citep{OBriain2021}. Neither a latent coordinate nor a domain mapping necessarily identifies the atomic input responsible for a feature.

We can instead use observed spectra to correct named physical inputs. Oscillator strengths set atomic line strengths, while damping parameters set their wings \citep{ValentiPiskunov1996,Gray2005}. These quantities combine laboratory measurements, theory, database evaluation, and calibration against benchmark stars \citep{Piskunov1995,Lobel2011,Jofre2014,Ryabchikova2015,Laverick2019,Smith2021}. A fast differentiable calculation allows many such inputs to be adjusted together while each correction remains tied to a physical parameter.

This distinction is important when a model misses the data. A flexible flux model may absorb the residual without deciding whether it came from an oscillator strength, a damping wing, the atmospheric structure, or a missing transition. Direct physical evaluation keeps these possibilities explicit and separately testable. It also allows a new abundance coordinate or atomic parameter to enter the governing calculation without regenerating a synthetic flux grid.

Direct physical inference therefore requires a much shorter calculation. An atmosphere calculation advances a depth-coupled state through sequential physical iterations. Once that structure is known, tens of thousands to millions of wavelength samples can be evaluated together. Established frameworks such as Kurucz, MARCS, PHOENIX, and TLUSTY make different physical and numerical choices, including LTE cool-star and non-LTE hot-star grids \citep{HubenyLanz1995,LanzHubeny2003,Kurucz2005,LanzHubeny2007, Gustafsson2008,Husser2013}. We focus on the original Kurucz programs, whose serial stages exchange intermediate files.

\paynezero\ reorganizes the supported one-dimensional LTE Kurucz physics for GPU synthesis and multicore CPU atmosphere iteration. Fixed data remain resident, and large compiled operations expose the independent work. The resulting forward model can be evaluated inside an optimizer.

The shorter runtime enables direct spectrum fitting and atomic-data calibration. After establishing synthesis, atmosphere iteration, and learned initialization, we first test a normalized-spectrum fitter whose final candidate is recalculated with a converged physical atmosphere. We then calibrate atomic line parameters against the Sun and Arcturus and combine both pieces with the instrumental response, velocity, broadening, continuum, and pixel uncertainties for the APOGEE application.

\section{Spectral Synthesis}\label{sec:synthesis}

Spectral synthesis turns a model atmosphere and a line list into an emergent spectrum. It calculates continuous and line opacity through the atmosphere, then solves radiative transfer to determine the radiation that leaves the stellar surface. \paynezero\ implements the supported Kurucz synthesis physics \citep{Kurucz2005} in a new wavelength-parallel calculation.

Once the depth state is fixed, different wavelengths have little dependence on one another. A broad spectrum can therefore replace one long wavelength loop with many copies of the same short depth calculation. This parallel structure is why the broad Kurucz synthesis benchmark below moves from more than ten minutes to seconds without an emulator. We begin with synthesis because it exposes the central computational idea directly. Before developing that calculation, Table~\ref{tab:atmosphere_symbols} collects the notation used throughout the paper.

\subsection{The physical calculation}

We sample the synthesis grid uniformly in logarithmic wavelength,
\begin{equation}
\lambda_{i_\lambda} =
\lambda_0\left(1+R_{\rm grid}^{-1}\right)^{i_\lambda} .
\end{equation}

\noindent The parameter $R_{\rm grid}$ sets the model sampling, not the instrumental resolving power. The instrument model applies the line spread function through $\mathcal{K}$, and several model samples can lie within one resolution element. The number of samples is approximately
\begin{equation}
N_\lambda \simeq
\frac{\ln(\lambda_{\max}/\lambda_{\min})}
 {\ln(1+R_{\rm grid}^{-1})}
\simeq R_{\rm grid}\ln\left(\frac{\lambda_{\max}}{\lambda_{\min}}\right).
\end{equation}

At each wavelength the mass extinction coefficient is the sum of continuous absorption, continuous scattering, and line absorption,
\begin{equation}
\chi_\nu = \kappa_{\nu,{\rm c}} + \sigma_{\nu,{\rm c}}
 + \sum_l \kappa_{\nu,l} .
\end{equation}

\noindent The continuous terms include the principal bound-free and free-free processes from hydrogen, helium, and metals \citep{Mihalas1978,Gray2005}. H$^-$ dominates much of the optical and near-infrared continuum in cool stars \citep{John1988}. Rayleigh and electron scattering remain separate because their source is not purely thermal.

For a standard atomic transition with a frequency normalized profile, the LTE line opacity per unit mass is \citep{Gray2005}
\begin{equation}
\begin{aligned}
\kappa_{\nu,l}={}&
\frac{\pi e^2}{m_e c}\,
\frac{n_{s,r}}{\rho U_{s,r}}\,(gf)_l
\exp\left(-\frac{E_l}{k_{\rm B}T}\right) \\
&{}\times
\left[1-\exp\left(-\frac{h\nu}{k_{\rm B}T}\right)\right]\phi_l(\nu), \\
&\text{with}\quad \int\phi_l(\nu)\,\mathrm{d}\nu=1 .
\end{aligned}
\end{equation}

\noindent Here $n_{s,r}/U_{s,r}$ is the partition-normalized number density of ion stage $r$ of species $s$, and $(gf)_l$ includes the lower-level statistical weight. The profile width and damping scale can be summarized by
\begin{equation}
\begin{aligned}
\Delta v_{\rm D}^{2} &=
\frac{2k_{\rm B}T}{m_l}+\xi^2, \\
\Delta\nu_{\rm D} &= \frac{\nu_l}{c}\Delta v_{\rm D}, \\
a_l &=
\frac{\gamma_{\rm rad}+\gamma_{\rm S}n_e+
\gamma_{\rm vdW}n_{\rm pert}}{4\pi\Delta\nu_{\rm D}} .
\end{aligned}
\end{equation}

\noindent Here $n_{\rm pert}$ is the effective neutral-collision density proxy used by the van der Waals prescription. The resulting Voigt profile sets how the central opacity is distributed in wavelength. At fixed population, the oscillator strength scales the frequency-integrated opacity, while damping redistributes opacity from the core into the wings \citep{Barklem2000,Gray2005}. Temperature, electron density, abundance, and ionization change both the number of absorbers and their broadening environment. These direct connections permit stellar labels and atomic data to be optimized in spectrum space.

The standard Voigt profile is not sufficient for every transition. We implement hydrogen fine structure, Stark profiles \citep{VidalCooperSmith1973}, and the merging of high series members into the continuum through occupation-probability physics \citep{HummerMihalas1988}. Separate Kurucz branches treat He~I, $^3$He, He~II, and asymmetric Shore--Fano autoionizing profiles \citep{Fano1961,Shore1967}. The specialized profiles follow \citet{KuruczAvrett1981}. Molecular line opacity uses populations from molecular equilibrium \citep{Tsuji1973} and molecular masses stored with the line data to set their Doppler widths.

The extinction defines optical depth through $\mathrm{d}\tau_\nu=\chi_\nu\,\mathrm{d}m$, where $m$ is column mass. In LTE the thermal source is the Planck function and line opacity is treated as thermal absorption. The continuum scattering source depends on the mean intensity \citep{Mihalas1978,Sobeck2011}. The source function can therefore be written as
\begin{equation}
\epsilon_\nu =
\frac{\kappa_{\nu,{\rm c}}+\sum_l\kappa_{\nu,l}}{\chi_\nu},
\qquad
S_\nu = \epsilon_\nu B_\nu(T)
 + \left(1-\epsilon_\nu\right)J_\nu .
\end{equation}

For each wavelength, after mapping the thermal source and scattering fraction to optical depth, we solve
\begin{equation}
\mu\frac{\mathrm{d}I_\nu}{\mathrm{d}\tau_\nu}=I_\nu-S_\nu .
\end{equation}

\noindent The transfer calculation iterates the scattering contribution. Since wavelengths remain independent, the full wavelength grid can be solved in parallel. With $\mu$ measured from the outward surface normal, the emergent Eddington flux is converted through
\begin{equation}
F_\lambda
=F_\nu\left|\frac{\mathrm{d}\nu}{\mathrm{d}\lambda}\right|
=4\pi H_\nu\frac{c}{\lambda^2}
=4\pi H_\lambda .
\end{equation}
\noindent The reported absolute flux is $F_\lambda$ per nanometer, with $c$ and $\lambda$ evaluated in consistent nanometer units. We carry two transfer branches for each wavelength. One contains line and continuum opacity. The other contains only the continuum. We solve them together to produce the total flux $F_\lambda$ and its physical continuum $F_{\lambda,{\rm c}}$. When an instrument operator is present, we apply it to both before taking their ratio,
\begin{equation}
f_{\rm norm} =
\frac{\mathcal{K}F_\lambda}
 {\mathcal{K}F_{\lambda,{\rm c}}} .
\end{equation}

\noindent This order matters whenever the continuum varies across a resolution element. It also lets an optimizer compare the model directly with observed pixels without moving a densely sampled spectrum back to CPU memory between synthesis and convolution. These equations define the physical calculation. The implementation changes how the same operations are arranged and executed.

\subsection{From staged files to wavelength-parallel execution}

\begin{table*}[t]
\centering
\caption{Symbols and definitions used for the atmosphere, synthesis, line-calibration, and spectral-inference calculations.}
\label{tab:atmosphere_symbols}
\begingroup
\setlength{\tabcolsep}{5pt}
\renewcommand{\arraystretch}{1.00}
\scriptsize
\begin{tabular}{@{}lll@{}}
\toprule
\tcell{0.14\textwidth}{\textbf{Role}} &
\tcell{0.25\textwidth}{\textbf{Symbol}} &
\tcell{0.55\textwidth}{\textbf{Definition}} \\
\midrule
\tcell{0.14\textwidth}{\textbf{Stellar labels}} &
\tcell{0.25\textwidth}{$T_{\rm eff}$, $\log g$, $g$, $\xi$,
$[\mathrm{M}/\mathrm{H}]$, $[\alpha/\mathrm{M}]$,
$[\mathrm{Fe}/\mathrm{H}]$, $\{[\mathrm{X}/\mathrm{H}]\}$,
$\{[\mathrm{X}/\mathrm{Fe}]\}$,
$\{[\mathrm{X}/\mathrm{M}]\}$, $A(\mathrm{X})$} &
\tcell{0.55\textwidth}{Effective temperature, logarithmic surface-gravity
label, physical acceleration $g=10^{\log g}$ in cgs, microturbulence, and
abundances. $[\mathrm{M}/\mathrm{H}]$ shifts all metals from the solar mixture,
and $[\alpha/\mathrm{M}]$ adds to O, Ne, Mg, Si, S, Ca, and Ti. Here
$[\mathrm{X}/\mathrm{H}]\equiv\log_{10}(n_{\rm X}/n_{\rm H})-
\log_{10}(n_{\rm X}/n_{\rm H})_\odot$,
$n_{\rm X}$ and $n_{\rm H}$ are elemental number densities, and the subscript
$\odot$ denotes the adopted solar reference.
$[\mathrm{Fe}/\mathrm{H}]$ is the same quantity for iron,
$[\mathrm{X}/\mathrm{Fe}]=[\mathrm{X}/\mathrm{H}]-
[\mathrm{Fe}/\mathrm{H}]$,
$[\mathrm{X}/\mathrm{M}]=[\mathrm{X}/\mathrm{H}]-[\mathrm{M}/\mathrm{H}]$,
and $A(\mathrm{X})=\log_{10}(n_{\rm X}/n_{\rm H})+12$. An explicit element
coordinate overrides bulk or group scaling} \\
\tcell{0.14\textwidth}{\textbf{Depth coordinate}} &
\tcell{0.25\textwidth}{$m$, $\tau_{\rm R}$} &
\tcell{0.55\textwidth}{Column mass increases inward. Rosseland optical depth
obeys $\mathrm{d}\tau_{\rm R}=\kappa_{\rm R}\,\mathrm{d}m$} \\
\tcell{0.14\textwidth}{\textbf{Atmosphere state}} &
\tcell{0.25\textwidth}{$T$, $\rho$, $P_{\rm gas}$, $n_e$, $\kappa_{\rm R}$,
$g_{\rm rad}$, $\xi$} &
\tcell{0.55\textwidth}{Temperature, mass density, gas pressure, electron
density, Rosseland opacity, and radiative acceleration at each depth, plus the
depth-independent microturbulent velocity} \\
\tcell{0.14\textwidth}{\textbf{Pressure terms}} &
\tcell{0.25\textwidth}{$P_{\rm rad}$, $P_0$} &
\tcell{0.55\textwidth}{Radiation pressure and the surface constant in
$P_{\rm gas}+P_{\rm rad}=gm+P_0$ for the branch without turbulent pressure} \\
\tcell{0.14\textwidth}{\textbf{Populations}} &
\tcell{0.25\textwidth}{$n_{s,r}$, $U_{s,r}$, $n_{\rm mol}$,
$n_{\rm pert}$, $\mathcal{I}_{s,r}$, $\Delta\mathcal{I}_{s,r}$} &
\tcell{0.55\textwidth}{Ion-stage number density and partition function for
species $s$ and ion stage $r$, molecular number density, neutral-collision
proxy, ionization energy, and its density-dependent lowering} \\
\tcell{0.14\textwidth}{\textbf{Energy balance}} &
\tcell{0.25\textwidth}{$F_{\rm rad}$, $F_{\rm conv}$, $\sigma_{\rm SB}$,
$\epsilon_F$} &
\tcell{0.55\textwidth}{Radiative and convective flux obey
$F_{\rm rad}+F_{\rm conv}=\sigma_{\rm SB}T_{\rm eff}^4$. The fractional
residual is $\epsilon_F$} \\
\tcell{0.14\textwidth}{\textbf{Iteration}} &
\tcell{0.25\textwidth}{$n$, $j$, $\Delta T^{(n)}$, $\epsilon_T$} &
\tcell{0.55\textwidth}{Iteration index, depth index, temperature correction,
and maximum deep layer relative temperature change} \\
\tcell{0.14\textwidth}{\textbf{Fixed point}} &
\tcell{0.25\textwidth}{$\mathbf{x}^{(n)}$, $\mathbf{p}^{(n)}$, $\mathcal{G}$,
 $\mathbf{J}_{\mathcal{G}}$, $\mathcal{R}$,
$\mathbf{x}_\star$, $\mathbf{e}^{(n)}$} &
\tcell{0.55\textwidth}{Numerical state and its six-profile projection,
one atmosphere cycle, local Jacobian and residual, converged fixed point and
iteration error} \\
\tcell{0.14\textwidth}{\textbf{Latent initializer}} &
\tcell{0.25\textwidth}{$\mathbf{u}$, $\overline{\mathbf u}$,
$\mathbf{s}_u$, $\widehat{\mathbf z}$, $\widehat{\mathbf c}$,
$\widehat{\mathbf u}$,
$\overline{\mathbf c}$, $\mathbf{s}_c$, $\mathbf{B}_K$, $K$, $\mathcal{D}$,
$\Delta m_j$, $T_{\rm grey}$, $s_g$} &
\tcell{0.55\textwidth}{Transformed profiles and their training mean and scale,
standardized network prediction, de-standardized predicted PCA coefficients,
decoded predicted transformed profiles, coefficient mean and scale, basis and
dimension, decoder, and mass, temperature, and acceleration transforms} \\
\tcell{0.14\textwidth}{\textbf{Training loss}} &
\tcell{0.25\textwidth}{$\mathcal{L}$, $\mathcal{L}_{\rm prof}$, $\mathcal{L}_{\nabla}$,
$\mathcal{L}_{\tau}$, $\mathcal{L}_{\rm hse}$,
$\lambda_{\nabla}$, $\lambda_{\tau}$, $\lambda_{\rm hse}$} &
\tcell{0.55\textwidth}{Total initializer objective, its decoded-profile,
depth-derivative, optical-depth, and hydrostatic terms, and their relative weights} \\
\tcell{0.14\textwidth}{\textbf{Reduced solver}} &
\tcell{0.25\textwidth}{$T(\tau_{\rm R})$, $m(\tau_{\rm R})$,
$\mathcal{R}_T$, $\mathcal{R}_m$} &
\tcell{0.55\textwidth}{Independent temperature and column mass profiles and
their energy balance and optical depth residuals} \\
\tcell{0.14\textwidth}{\textbf{Constants}} &
\tcell{0.25\textwidth}{$c$, $h$, $k_{\rm B}$, $e$, $m_e$} &
\tcell{0.55\textwidth}{Speed of light, Planck and Boltzmann constants,
elementary charge, and electron mass} \\
\tcell{0.14\textwidth}{\textbf{Atomic line}} &
\tcell{0.25\textwidth}{$l$, $\nu_l$, $(gf)_l$, $E_l$, $m_l$} &
\tcell{0.55\textwidth}{Transition index, frequency, statistically weighted
oscillator strength, lower-level energy, and absorber mass} \\
\tcell{0.14\textwidth}{\textbf{Broadening}} &
\tcell{0.25\textwidth}{$\gamma_{\rm rad}$, $\gamma_{\rm S}$,
$\gamma_{\rm vdW}$} &
\tcell{0.55\textwidth}{Radiative damping rate and per-perturber Stark and van
der Waals coefficients, which multiply the electron and neutral-collision
densities} \\
\tcell{0.14\textwidth}{\textbf{Line calibration}} &
\tcell{0.25\textwidth}{$q(l)$, $\beta$, $\delta_{q,gf}$,
$\delta_{q,\beta}$} &
\tcell{0.55\textwidth}{Correction group assigned to line component $l$,
damping-family index, and bounded corrections to oscillator strength and
damping. A prime denotes a calibrated component value} \\
\tcell{0.14\textwidth}{\textbf{Sampling}} &
\tcell{0.25\textwidth}{$\nu$, $\lambda_{i_\lambda}$, $\lambda_0$,
$\lambda_{\min}$, $\lambda_{\max}$, $R_{\rm grid}$, $N_\lambda$} &
\tcell{0.55\textwidth}{Frequency, wavelength coordinate, first sample, and
requested endpoints. At synthesis index $i_\lambda$, the grid obeys
$\lambda_{i_\lambda+1}/\lambda_{i_\lambda}=1+R_{\rm grid}^{-1}$ and contains
$N_\lambda$ samples} \\
\tcell{0.14\textwidth}{\textbf{Profile}} &
\tcell{0.25\textwidth}{$\Delta v_{\rm D}$, $\Delta\nu_{\rm D}$, $a_l$,
$\phi_l$} &
\tcell{0.55\textwidth}{Doppler velocity and frequency widths, damping ratio,
and line profile normalized by $\int\phi_l(\nu)\,\mathrm{d}\nu=1$} \\
\tcell{0.14\textwidth}{\textbf{Opacity}} &
\tcell{0.25\textwidth}{$\kappa_{\nu,{\rm c}}$, $\sigma_{\nu,{\rm c}}$,
$\kappa_{\nu,l}$, $\chi_\nu$} &
\tcell{0.55\textwidth}{Continuous absorption, continuous scattering, and line
absorption combine as $\chi_\nu=\kappa_{\nu,{\rm c}}+
\sigma_{\nu,{\rm c}}+\sum_l\kappa_{\nu,l}$} \\
\tcell{0.14\textwidth}{\textbf{Line selection}} &
\tcell{0.25\textwidth}{$b$, $\nu_b$, $\eta_{\rm sel}$, $C_{jb}$, $Q_{srb}$, $C_0$,
$R_l$, $T_*$} &
\tcell{0.55\textwidth}{Frequency-bin index and its reference frequency,
selection fraction, continuum threshold, maximum population-to-continuum
proxy, fixed conversion constant, and central line-strength proxy. $T_*$ is
the deepest-layer temperature used by the conservative keep test} \\
\addlinespace[2pt]
\tcell{0.14\textwidth}{\textbf{Transfer}} &
\tcell{0.25\textwidth}{$\tau_\nu$, $\epsilon_\nu$, $\mu$, $B_\nu$, $J_\nu$,
$S_\nu$, $I_\nu$} &
\tcell{0.55\textwidth}{Optical depth, thermal absorption fraction, outward
direction cosine, Planck function, mean intensity, source function, and
specific intensity} \\
\tcell{0.14\textwidth}{\textbf{Output}} &
\tcell{0.25\textwidth}{$H_\nu$, $H_\lambda$, $F_\nu$, $F_\lambda$,
$F_{\lambda,{\rm c}}$, $f_{\rm norm}$} &
\tcell{0.55\textwidth}{$H_\nu$ and $H_\lambda$ are first angular moments, with
$F_\nu=4\pi H_\nu$ and $F_\lambda=4\pi H_\lambda$. $F_\nu(m)$ is the
monochromatic flux within the atmosphere. Its surface value and the emergent
$F_\lambda$ are the spectrum, $F_{\lambda,{\rm c}}$ its continuum, and
$f_{\rm norm}$ their ratio} \\
\tcell{0.14\textwidth}{\textbf{Instrument}} &
\tcell{0.25\textwidth}{$\mathcal{K}$} &
\tcell{0.55\textwidth}{Linear line-spread-function operator from intrinsic
total and continuum flux to the observed grid} \\
\tcell{0.14\textwidth}{\textbf{Spectrum fit}} &
\tcell{0.25\textwidth}{$\boldsymbol{\theta}$, $y_p$, $w_p$, $f_p$,
$f_{\rm fast}$, $f_{\rm phys}$, $\Delta f$, $Q(f)$, $\delta Q_{\min}$} &
\tcell{0.55\textwidth}{Fitted stellar labels, observed normalized flux,
inverse-variance weight, and modeled normalized flux at pixel $p$.
$f_{\rm fast}$ and $f_{\rm phys}$ are the spectra before and after atmosphere
convergence at the fast minimum, $\Delta f=f_{\rm phys}-f_{\rm fast}$,
$Q$ is the mean chi-square per valid pixel, and $\delta Q_{\min}$ is the minimum
predicted decrease required before another physical atmosphere calculation} \\
\tcell{0.14\textwidth}{\textbf{Survey nuisance}} &
\tcell{0.25\textwidth}{$\boldsymbol{\psi}$, $v_r$, $v_b$, $\mathcal{S}$, $\mathcal{B}$} &
\tcell{0.55\textwidth}{Joint survey-fit state, residual velocity,
effective Gaussian broadening,
Doppler operator, and broadening operator} \\
\addlinespace[2pt]
\bottomrule
\end{tabular}
\endgroup
\end{table*}

\begin{table*}[t]
\centering
\caption{Execution of the same synthesis stages in the original Kurucz workflow and \paynezero. The final column summarizes how resident data and wavelength-parallel GPU kernels change the computational cost.}
\label{tab:synthesis_speed}
\begingroup
\setlength{\tabcolsep}{4pt}
\renewcommand{\arraystretch}{1.18}
\compacttablefont
\begin{tabular}{@{}llll@{}}
\toprule
\tcell{0.12\textwidth}{\textbf{Stage}} &
\tcell{0.25\textwidth}{\textbf{Original Kurucz execution}} &
\tcell{0.28\textwidth}{\textbf{\paynezero\ execution}} &
\tcell{0.25\textwidth}{\textbf{Effect}} \\
\midrule
\tcell{0.12\textwidth}{\textbf{Window preparation}} &
\tcell{0.25\textwidth}{Line readers prepare wavelength-specific sequential
records for the staged programs} &
\tcell{0.28\textwidth}{Numba compiles molecular source records into portable
window-bound arrays. An in-process cache retains the grid, catalogs, tables,
and fixed tensors for one device, dtype, and chunk configuration} &
\tcell{0.25\textwidth}{The retained measurements of repeated calls amortize this
preparation without making the disk products device specific} \\
\tcell{0.12\textwidth}{\textbf{Work layout}} &
\tcell{0.25\textwidth}{Compiled Fortran routines retain serial outer loops over
depth records and wavelengths} &
\tcell{0.28\textwidth}{PyTorch batches independent wavelengths and the dominant
narrow-line and depth contributions. The retained V100, A100, and H100
measurements use CUDA. The same path also supports Metal and CPU execution} &
\tcell{0.25\textwidth}{The bulk work advances concurrently, while Python dispatch is
limited to bounded chunks and specialized profiles} \\
\tcell{0.12\textwidth}{\textbf{Line opacity}} &
\tcell{0.25\textwidth}{Each depth rereads atomic and molecular line records and
advances through profiles in record order} &
\tcell{0.28\textwidth}{Atomic and molecular catalogs are processed in bounded
line chunks. A continuum-referenced center test rejects weak depth--line pairs,
and survivors deposit only their required wing spans. Both tests use the
same fraction of the local continuum opacity, $10^{-3}$, as the original Fortran} &
\tcell{0.25\textwidth}{Cost and temporary memory follow retained-line scans and
active profile deposits, not the full line-list--wavelength product} \\
\tcell{0.12\textwidth}{\textbf{Memory and I/O}} &
\tcell{0.25\textwidth}{Depth-ordered opacity is written and transposed through
intermediate files} &
\tcell{0.28\textwidth}{Depth--wavelength opacity, source, and transfer arrays
stay on the device between stages, while line and active-pair chunks bound temporary
memory} &
\tcell{0.25\textwidth}{Disk opacity staging and transfers of full arrays through CPU memory are
removed without constructing a depth--line--wavelength array} \\
\tcell{0.12\textwidth}{\textbf{Transfer}} &
\tcell{0.25\textwidth}{Wavelengths advance serially, with separate total and
continuum calls} &
\tcell{0.28\textwidth}{All wavelengths and the stacked total and continuum
branches share one tensor calculation} &
\tcell{0.25\textwidth}{The largest independent dimension is evaluated
together} \\
\addlinespace[2pt]
\bottomrule
\end{tabular}
\endgroup
\end{table*}

The original Kurucz calculation runs as a sequence of Fortran programs. Line readers first prepare a requested wavelength interval. The opacity stage then loops over atmosphere depth, rewinds the line records for each layer, accumulates line opacity, and writes depth-ordered records. Direct-access files transpose these records. The transfer stage reads the transposed opacity and advances through wavelength one point at a time. At every point it solves the continuum and total transfer problems. This organization was well matched to the memory available when the programs were written. It also creates intermediate input and output and leaves the largest independent dimension largely serial \citep{Kurucz2005}.

The opportunity for parallel execution follows from the physical dependencies. Radiative transfer couples the atmosphere depths at one wavelength, but it does not couple one wavelength to the next. The natural unit of work is therefore a large set of independent wavelength columns, each with a short calculation through depth. The historical depth-ordered records and file transpose obscure this structure. Keeping the complete depth and wavelength array in memory exposes it directly.

A GPU is effective when one compiled operation acts on many array elements at once. Such an operation is often called a kernel. In \paynezero, we pass each kernel a large group of wavelengths or active lines rather than one wavelength. We leave the physical equations unchanged, but repeated small calls become compiled array operations. Continuum opacity spans the full depth and wavelength array, active lines add their profiles in parallel, and we integrate optical depth along the short depth axis for all wavelengths. Total and continuum transfer share the same batch. We keep the arrays and fixed transfer geometry on the GPU until the kernel returns the final flux.

\begin{figure*}[t]
\centering
\includegraphics[width=\textwidth]{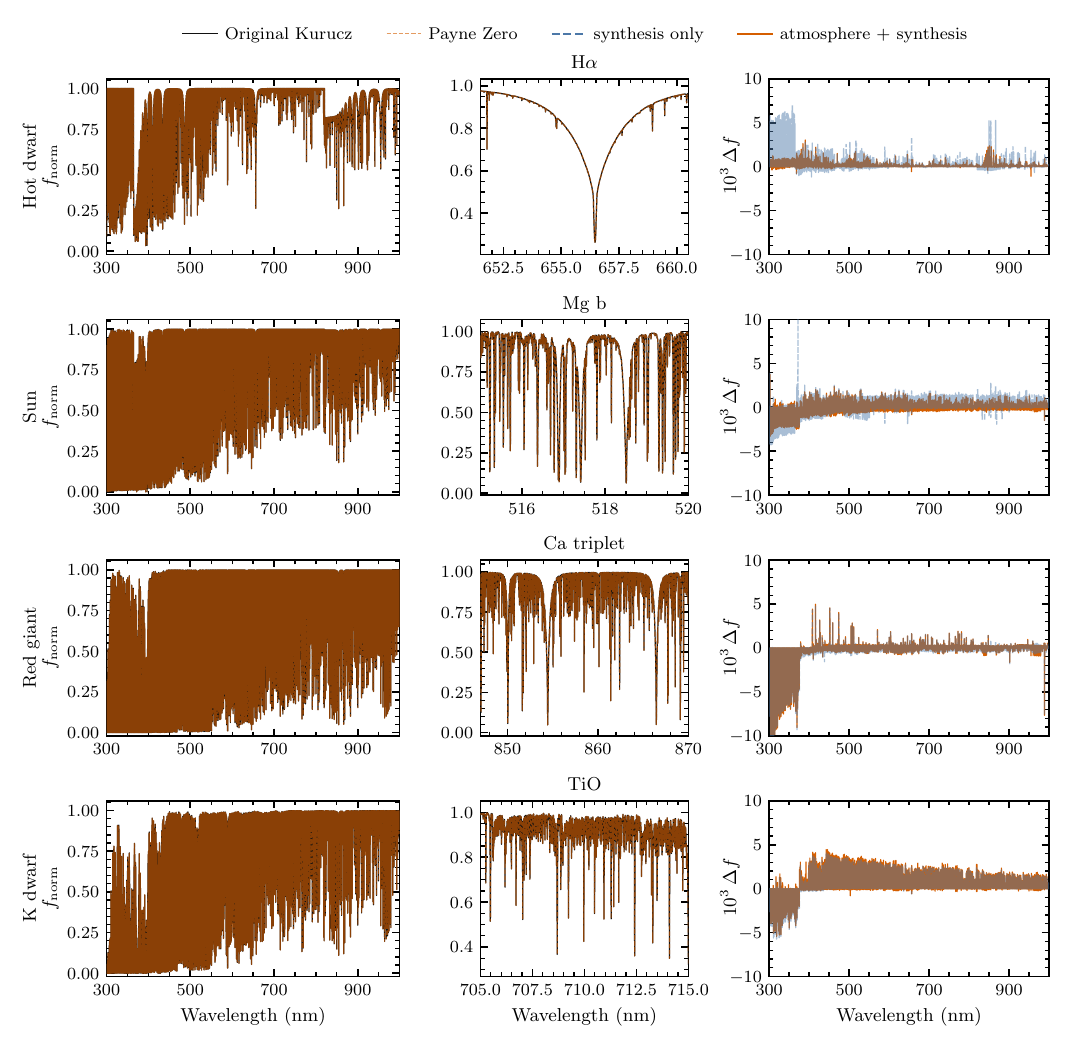}
\caption{Agreement with the original Kurucz calculation over 300--1000 nm at $R_{\rm grid}=300{,}000$. Rows show a hot dwarf, the Sun, a red giant, and a K dwarf. Columns show the full normalized spectrum, a representative feature, and $\Delta f_{\rm norm}=f_{\rm norm}^{\mathrm{Payne\,Zero}}- f_{\rm norm}^{\rm Kurucz}$. Black and orange compare the complete calculations. In the residual panels, blue holds the atmosphere fixed to isolate synthesis, while orange includes independently converged atmospheres at the same labels. Residuals are multiplied by $10^3$. From top to bottom, the synthesis-only and complete 99th-percentile absolute residuals are, in units of $10^{-3}$, $(2.0,0.7)$, $(2.8,2.2)$, $(5.8,5.8)$, and $(3.0,2.9)$.}
\label{fig:synthesis_overview}
\end{figure*}

Two thresholds control this line work, and both carry the values used by
the original programs. A depth--line pair is deposited only when its central
line opacity reaches $10^{-3}$ of the local continuum opacity, and the wings of a
retained line are traced outward until the profile falls below the same
$10^{-3}$ fraction. The original Fortran applies one parameter in the same two places and the retained benchmarks set it to
$10^{-3}$, so the physical selection is unchanged rather than newly imposed.

Because one fraction sets both the center test and the retained wing
extent, lowering it admits weaker lines and traces retained wings further at the
same time. Taking it from $10^{-3}$ to $10^{-6}$ for the Sun and a red giant
changes the normalized flux by 2.5 and $3.0\times10^{-3}$ rms. The change
decelerates over that range without fully converging, and it is about three
times the residual against the original Kurucz calculation reported in
Section~\ref{sec:synthesis}. Both calculations apply the same fraction, so the
retained transitions are the same in each and that comparison is unaffected. The size of the effect is a property
of the shared approximation rather than of either implementation.

Repeating the standard-star calibration of Section~\ref{sec:linecal} with
this fraction tightened by two decades leaves the fitted corrections slightly
larger rather than smaller. The typical oscillator-strength correction moves from
0.070 to 0.079~dex and the typical van der Waals correction from 0.040 to
0.046~dex, with more transitions adjusted as more lines enter the fit. The fitted
damping is therefore not compensating for truncated wings. The corrections
themselves shift by about 13 percent between the two thresholds, so the released
values belong with the fraction they were fitted at.

The remaining bounds partition the work rather than the physics. Lines are grouped into reach tiers so that
short-winged transitions are deposited by a batched scatter while the few
long-winged transitions fall back to a per-line walk, which partitions the work
without truncating any profile, and a cap of $10^{6}$ profile steps guards
against an unbounded walk and is not reached at the sampling used here.
The dominant narrow-line and wavelength calculations therefore avoid Python loops over individual profiles or wavelength samples. Specialized broad and autoionizing profiles and bounded catalog chunks retain small Python-level dispatch loops. PyTorch sends their large array operations to CUDA or CPU kernels, while a cached Numba kernel prepares the molecular catalog. The tested Apple Metal path makes synthesis available locally on Apple silicon. The reported timings and throughput recommendations use dedicated NVIDIA V100, A100, and H100 accelerators. Since the Kurucz programs are also compiled, the speed difference comes from concurrency, larger kernels, and resident inputs rather than compilation alone. Table~\ref{tab:synthesis_speed} summarizes how these changes replace serial loops and intermediate files. During repeated fits, the fixed window, catalog, and transfer geometry remain resident, while atmosphere-dependent opacity and flux are recomputed for every label vector. No synthetic flux is cached or emulated.

\subsection{Accuracy and performance}
\label{sec:accuracy_performance}

We test the final spectrum over 300--1000~nm on a logarithmic grid with $R_{\rm grid}=300{,}000$. Figure~\ref{fig:synthesis_overview} makes two comparisons. A shared atmosphere and line catalog isolate synthesis accuracy. Independent calculations at the same stellar labels test atmosphere and synthesis together. Both use the original Fortran Kurucz chain as the reference.\footnote{\url{https://github.com/tingyuansen/kurucz}}

Across the four stellar types, the pooled rms normalized-flux residual is $9.1\times10^{-4}$ for synthesis alone and $8.6\times10^{-4}$ for the complete calculation. For the hot dwarf, Sun, red giant, and K dwarf, respectively, the 99th-percentile absolute residuals are $(2.0,2.8,5.8,3.0)\times10^{-3}$ for synthesis alone and $(0.7,2.2,5.8,2.9)\times10^{-3}$ for the complete calculation. These residuals reflect small numerical differences between two implementations of the same physical calculation.

\begin{figure*}[t]
\centering
\includegraphics[width=\textwidth]{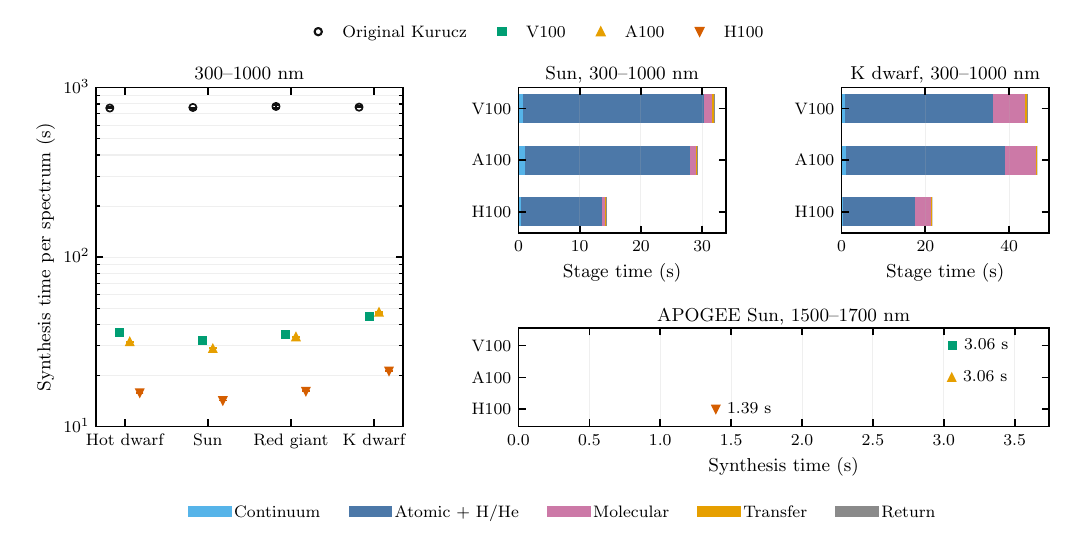}
\caption{Synthesis runtime per spectrum at $R_{\rm grid}=300{,}000$. The left panel compares 300--1000 nm calculations for four stellar types with the original serial Kurucz programs and \paynezero\ on one NVIDIA V100, A100, or H100 GPU. The upper middle and right panels decompose the GPU wall for the Sun and K dwarf. The lower panel gives the total solar wall over 1500--1700 nm. Atomic and molecular line opacity dominate the GPU runtime.}
\label{fig:synthesis_runtime}
\end{figure*}

Figure~\ref{fig:synthesis_runtime} measures repeated forward calculations after one-time source-window preparation. Each wall includes continuum opacity, atomic and molecular lines, total and continuum transfer, and the returned spectrum. On the 300--1000 nm grid at $R_{\rm grid}=300{,}000$, the solar medians are about 763~s for the original Kurucz programs, 32~s for one V100, 29~s for one A100, and 14~s for one H100. These correspond to speedups of $24$, $26$, and $54$ on the V100, A100, and H100, respectively.

For the same solar atmosphere over the narrower APOGEE 1500--1700~nm interval, the V100, A100, and H100 medians are about 3.1, 3.1, and 1.4~s. All use $R_{\rm grid}=300{,}000$ and are native-grid synthesis timings before any instrumental convolution or resampling. The APOGEE benchmark changes only the wavelength interval, not the synthesis resolution.

Two costs are paid once for a process rather than once per spectrum.
Importing the package and creating the device context together take about one
second. Building the cached window products is paid once for a given
window and takes seconds to a minute depending on the host. Once those products exist, a new process
reaches its first spectrum in under one second.

Peak device memory over the 1500--1700~nm window at
$R_{\rm grid}=300{,}000$ is 8.2~GB, and it is the same on both accelerators and
in both precisions. Memory follows the depth--wavelength arrays and the bounded
line chunks rather than the device, so a large-memory accelerator is not
required.
The V100 runs use the V100S variant and 2,000-line chunks, while the A100 and H100 runs use 4,000- and 8,000-line chunks. The V100 and A100 walls are close because atomic and H/He opacity accounts for 80--92 percent of the profiled Sun and K-dwarf calculations, and their measured times for this dominant stage differ by less than 10 percent. The H100 roughly halves the same stage. These profile-evaluation and indexed-accumulation kernels do not use dense tensor-core matrix multiplication. The GPUs ran in their available host, PyTorch, and CUDA environments. These are therefore achieved application timings rather than an isolated architecture comparison.

Synthesis carries float64 by default and float32 on Metal, which has no
float64 path. We repeated one solar 1500--1700~nm spectrum while varying one
condition at a time. Repeating the calculation on a fixed device, dtype, and
chunk size reproduces the spectrum bitwise on both an A100 and an H100, so the
indexed line accumulation is deterministic within a device. Changing the
molecular chunk size alters at most 600 of 37,549 samples by $1.0\times10^{-8}$,
and moving between the two accelerators alters the spectrum by at most
$1.4\times10^{-8}$. Evaluating in float32 rather than float64 shifts every sample,
by at most $2.6\times10^{-6}$.
Even the largest of these is more than two orders of magnitude below the
$9.1\times10^{-4}$ rms difference from the original Kurucz calculation, so device,
precision, and chunking do not affect the comparisons reported here.

Synthesis time depends only weakly on grid sampling. Over the fixed 300--1000 nm interval, increasing $R_{\rm grid}$ by a factor of 30 raises the wall by factors of only 3.1--5.0 across the four stars. Denser sampling adds transfer columns and pixels within each active profile, while scanning the retained catalog and the fixed per-call floor remain nearly unchanged.

\begin{figure}[t]
\centering
\includegraphics[width=\columnwidth]{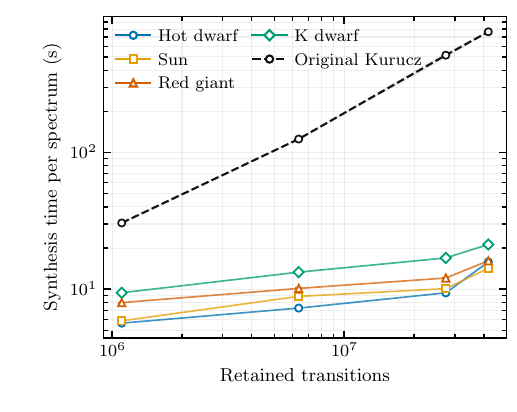}
\caption{Synthesis scaling with wavelength span at $R_{\rm grid}=300{,}000$. All windows begin at 300~nm and end at 400, 500, 700, or 1000~nm. Colored curves show the four stellar types on one H100. The dashed curve and shaded band give the median and range of the original Kurucz times for the same stars. The shallower H100 scaling increases the speedup as the spectral interval and retained transition count grow.}
\label{fig:synthesis_scaling}
\end{figure}

Figure~\ref{fig:synthesis_scaling} instead tests wavelength span. At fixed $R_{\rm grid}=300{,}000$, expanding the window from a 100 to a 400~nm span increases the retained catalog by a factor of 25. The H100 wall rises by only factors of 1.5--1.8, whereas the original Kurucz wall rises by factors of 16--18. The smaller GPU gain for the shortest window reflects less work to fill the accelerator. A fixed per-call floor is amortized as more wavelengths and profiles enter the batch. Moreover, the retained count is only an upper bound on opacity work because the continuum cutoff rejects most depth--line pairs and surviving profiles visit bounded pixel ranges. The H100 speedup therefore grows from about $3$--$6$ for the 100~nm span to $36$--$54$ over 300--1000~nm.

\section{Model Atmospheres}\label{sec:atmosphere}

The atmosphere calculation supplies the depth structure used by synthesis. At specified effective temperature, surface gravity, composition, and microturbulence, it makes the pressure, populations, opacity, radiation field, and convection mutually consistent. \paynezero\ implements the one-dimensional, plane-parallel LTE Kurucz atmosphere calculation used here \citep{Kurucz2005} with compiled array kernels. We do not include optional NLTE or turbulent-pressure branches.

Atmospheric structure must be solved iteratively because its variables form a feedback loop. Temperature and pressure set the populations, the populations set the opacity, and the opacity sets the radiative flux and heating. The resulting flux imbalance changes the temperature and depth structure used by the next iteration. Its computational cost is therefore set by both the work in each physical pass and the number of passes needed to reach a consistent structure.

\subsection{The physical calculation}

The production grid contains 80 layers in Rosseland optical depth. We summarize its physical structure with six principal profiles. These are column mass $m$, temperature $T$, gas pressure $P_{\rm gas}$, electron density $n_e$, Rosseland opacity $\kappa_{\rm R}$, and radiative acceleration $g_{\rm rad}$. The iteration also carries the radiation-pressure and convection state needed to update these profiles. Microturbulence $\xi$ is fixed by the requested labels. The converged product includes the populations and Doppler widths needed by synthesis.

Hydrostatic equilibrium relates pressure to the weight of material above each layer. In the integrated plane parallel form used here,
\begin{equation}
P_{\rm gas}(m)+P_{\rm rad}(m)
=g m+P_0 .
\end{equation}
\noindent Here $P_{\rm rad}$ is the radiation pressure and $P_0$ is the surface integration constant. This is the convention for the supported branch, in which turbulent pressure is disabled. The equation of state then solves charge conservation, Saha ionization, Boltzmann populations, and molecular equilibrium at each layer \citep{Tsuji1973,Mihalas1978,HubenyMihalas2014}. Adjacent ion stages obey
\begin{equation}
\begin{aligned}
\frac{n_{s,r+1}n_e}{n_{s,r}}
&=\frac{2U_{s,r+1}}{U_{s,r}}
\left(\frac{2\pi m_e k_{\rm B}T}{h^2}\right)^{3/2} \\
&\quad\times
\exp\left[-\frac{\mathcal{I}_{s,r}-\Delta\mathcal{I}_{s,r}}
{k_{\rm B}T}\right].
\end{aligned}
\end{equation}
\noindent Here $\Delta\mathcal{I}_{s,r}$ is the density-dependent lowering of the isolated ionization energy in the adopted prescription. Its partition functions include the corresponding occupation correction. Charge conservation closes the electron density, while molecular mass action closes the molecular populations \citep{Tsuji1973,BarklemCollet2016}. Together these calculations produce the density and populations used by the opacity calculation.

\begin{table*}[t]
\centering
\caption{Execution of the same atmosphere stages in the original Kurucz workflow and \paynezero. The final column summarizes how resident data, compiled kernels, and multicore frequency parallelism change the computational cost.}
\label{tab:atmosphere_speed}
\begingroup
\setlength{\tabcolsep}{4pt}
\renewcommand{\arraystretch}{1.18}
\compacttablefont
\begin{tabular}{@{}llll@{}}
\toprule
\tcell{0.12\textwidth}{\textbf{Stage}} &
\tcell{0.25\textwidth}{\textbf{Original Kurucz execution}} &
\tcell{0.28\textwidth}{\textbf{\paynezero\ execution}} &
\tcell{0.25\textwidth}{\textbf{Effect}} \\
\midrule
\tcell{0.12\textwidth}{\textbf{Physical inputs}} &
\tcell{0.25\textwidth}{Input decks and line catalogs use fixed-width or
Fortran-unformatted records. Many numerical tables are compiled into the source} &
\tcell{0.28\textwidth}{Most fixed physics tables and the full line-source
catalogs are retained as typed in-process arrays} &
\tcell{0.25\textwidth}{Cached tables and line catalogs avoid repeated parsing,
although isotope and molecular-equilibrium inputs are still reloaded each pass} \\
\tcell{0.12\textwidth}{\textbf{Line selection}} &
\tcell{0.25\textwidth}{A separate first-pass scan writes the
atmosphere-dependent selected-line file} &
\tcell{0.28\textwidth}{Once per atmosphere, fused parallel keep tests over each
catalog family reuse cached columns and bin assignments and return the selected
catalog in memory. Later passes reuse it} &
\tcell{0.25\textwidth}{Selection remains atmosphere dependent, without a
selected-line file round trip} \\
\tcell{0.12\textwidth}{\textbf{Populations and continuum}} &
\tcell{0.25\textwidth}{Population and continuum calculations advance through
serial nested depth, species, and frequency loops} &
\tcell{0.28\textwidth}{Population and molecular-equilibrium kernels are compiled
but depth-ordered, while continuum kernels parallelize over sampling frequency} &
\tcell{0.25\textwidth}{Continuum frequency work uses the configured CPU threads,
whereas population work remains ordered} \\
\tcell{0.12\textwidth}{\textbf{Line opacity}} &
\tcell{0.25\textwidth}{Selected and detailed line records are scanned serially
and deposited through nested depth and wavelength loops} &
\tcell{0.28\textwidth}{Compiled parallel chunks accumulate selected and
detailed transitions on the opacity-sampling grid} &
\tcell{0.25\textwidth}{The dominant line-opacity work uses the configured CPU
threads} \\
\tcell{0.12\textwidth}{\textbf{Transfer}} &
\tcell{0.25\textwidth}{Each of 30,000 wavelengths is solved and integrated
before the next} &
\tcell{0.28\textwidth}{Compiled frequency chunks solve transfer in parallel into
private accumulators, which are reduced after the parallel region} &
\tcell{0.25\textwidth}{Independent frequency work is parallel, while the
outer physical cycle remains ordered} \\
\addlinespace[2pt]
\bottomrule
\end{tabular}
\endgroup
\end{table*}

The atmosphere solver evaluates line blanketing by direct opacity sampling, a method that represents large atomic and molecular line ensembles at selected frequencies \citep{JohnsonKrupp1976}. For each requested abundance mixture, the solver recomputes the equation of state and every represented atomic and molecular opacity \citep{Kurucz1996,KuruczAtlas12ASCL}. These internal wavelength samples determine radiative heating and flux during atmosphere iteration.

At each sampling wavelength, the solver evaluates continuous and line extinction through all depth layers \citep{Mihalas1978}. The Rosseland mean gives the opacity relevant to diffusive energy transport,
\begin{equation}
\frac{1}{\kappa_{\rm R}}=
\frac{\displaystyle\int_0^\infty
 \chi_\nu^{-1}\frac{\partial B_\nu}{\partial T}\,\mathrm{d}\nu}
 {\displaystyle\int_0^\infty
 \frac{\partial B_\nu}{\partial T}\,\mathrm{d}\nu},
\qquad \mathrm{d}\tau_{\rm R}=\kappa_{\rm R}\,\mathrm{d}m .
\end{equation}
\noindent Here $\chi_\nu$ is the total mass extinction. Radiative transfer at the same sampling wavelengths supplies the monochromatic flux, radiation pressure, radiative acceleration, and heating integrals. In particular, the acceleration from the radiation field is
\begin{equation}
g_{\rm rad}(m)=\frac{1}{c}\int_0^\infty
\chi_\nu(m)F_\nu(m)\,\mathrm{d}\nu .
\end{equation}
\noindent In layers where convection is active, a mixing-length calculation supplies the convective flux \citep{BohmVitense1958,Mihalas1978,Gustafsson2008}. The target energy balance and its fractional residual are
\begin{equation}
\begin{aligned}
F_{\rm rad}(m)+F_{\rm conv}(m)
&=\sigma_{\rm SB}T_{\rm eff}^{4}, \\
\epsilon_F(m)
&=\frac{F_{\rm rad}(m)+F_{\rm conv}(m)}
 {\sigma_{\rm SB}T_{\rm eff}^{4}}-1 .
\end{aligned}
\end{equation}

\noindent Let $\mathbf{x}^{(n)}$ denote the full remapped atmosphere state entering pass $n$. Its principal profiles are $m$, $T$, $P_{\rm gas}$, $n_e$, $\kappa_{\rm R}$, and $g_{\rm rad}$. It also carries the pressure and convection quantities needed by the next pass. One physical solver iteration then has the schematic dependence
\begin{equation}
\begin{aligned}
\mathbf{x}^{(n)}
&\longrightarrow
\left(P_{\rm gas},n_e,\{n_{s,r}\},\{n_{\rm mol}\}\right)^{(n)} \\
&\longrightarrow
\chi_\nu^{(n)} \longrightarrow
\left(J_\nu,F_{\rm rad},g_{\rm rad},\kappa_{\rm R},\tau_{\rm R}\right)^{(n)} \\
&\longrightarrow \left(F_{\rm conv},\Delta T\right)^{(n)}
\longrightarrow \left(T,m\right)^{(n+1)} \\
&\xrightarrow{\ \rm remap\ } \mathbf{x}^{(n+1)},
\qquad T^{(n+1)}=T^{(n)}+\Delta T^{(n)} .
\end{aligned}
\end{equation}
\noindent Thus hydrostatic balance and the equation of state set pressure and populations before the opacity and transfer calculation determines the fluxes, radiative acceleration, and temperature correction. The correction also updates column mass. Both corrected profiles, together with the other carried fields, are remapped to the standard Rosseland grid. The next pass recomputes pressure and every population from this remapped state.

We adapt the deep-layer successive-structure norm used by the external checker in the retained Kurucz workflow into a per-pass production stopping test,
\begin{equation}
\label{eq:atmosphere_stop}
\epsilon_T=\max_{j\in {\rm deep}}
\left|\frac{T_j^{(n+1)}-T_j^{(n)}}{T_j^{(n+1)}}\right|
<5\times10^{-4} .
\end{equation}
\noindent Here ${\rm deep}=\{40,\ldots,75\}$ on the 80-layer production grid. This norm tests whether the deep structure has ceased changing appreciably. Because it is local, we check flux balance and the emergent spectrum separately.

\subsection{From serial iteration to parallel CPU execution}

The original Kurucz atmosphere program carries this physical cycle through a serial outer iteration. Within each iteration it loops through the depth layers to update pressure and populations, constructs line opacity, and then advances through the 30,000 opacity sampling wavelengths. At each wavelength it evaluates the continuous opacity, solves radiative transfer through all layers, and adds the result to the Rosseland, radiation pressure, and temperature correction integrals. Only after this frequency loop is complete can convection and the next temperature structure be calculated.

Successive iterations cannot overlap because the temperature correction from pass $n$ changes the populations and opacity of pass $n+1$. Within one pass, however, work over depth, species, transitions, and opacity-sampling wavelengths is independent until the partial results are reduced. \paynezero\ parallelizes these blocks, reuses fixed inputs, and removes repeated parsing and staging.

\paynezero\ keeps the physical ordering and correction equations of the outer atmosphere cycle. In the retained original-program benchmark, the solver executes the requested number of iterations exactly. Each batch contains 30 iterations. A separate checker then compares the last two structures and launches another batch when needed. \paynezero\ instead evaluates Equation~\ref{eq:atmosphere_stop} after every pass following a three-pass minimum, stopping at the threshold or the configured cap.

Within each pass, the performance-critical loops are written in Python. Numba's just-in-time compiler turns functions marked \texttt{njit} into machine code, while \texttt{prange} divides independent loop iterations among CPU threads. In the molecules-enabled production path, population and molecular-equilibrium kernels remain depth-ordered, while continuum-frequency, line-opacity, and transfer work use this threaded execution. Most fixed physics tables and the full line-source catalogs remain in typed process arrays. Since the reference solver is already compiled Fortran, JIT compilation alone does not explain the gain. Larger kernels, resident inputs, and removal of intermediate file staging supply the remaining reduction.

\begin{figure*}[t]
\centering
\includegraphics[width=\textwidth]{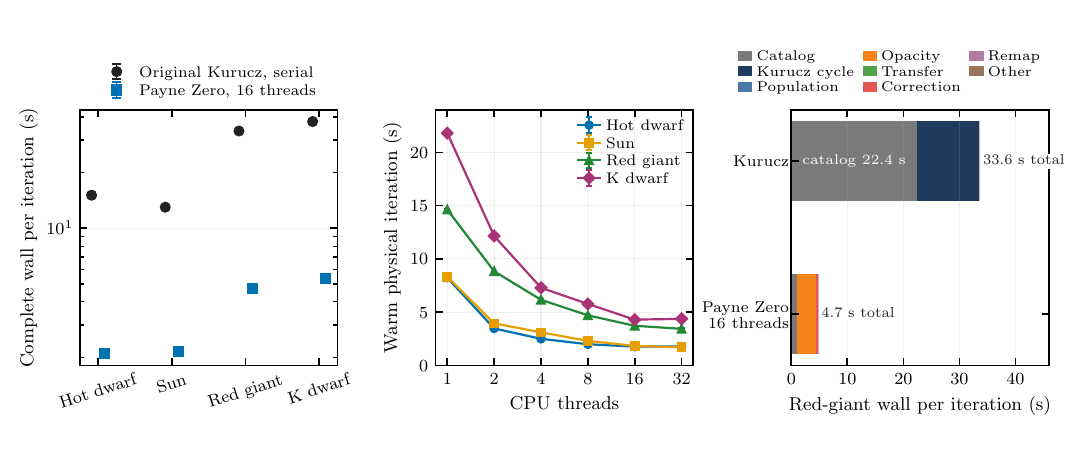}
\caption{Atmosphere performance on an AMD EPYC 9B45 CPU. The left panel compares the mean wall per physical iteration from matched three-iteration runs for four stellar types. \paynezero\ uses 16 threads. The middle panel shows its iteration wall from 1 to 32 threads. The right panel decomposes the red-giant iteration into physical stages for original Kurucz and 16-thread \paynezero. The latter is 6--7 times faster across the four stars, with line opacity the largest remaining stage.}
\label{fig:atmosphere_performance}
\end{figure*}

Line selection illustrates these changes. The source catalog is fixed, but the subset retained for opacity sampling depends on the atmosphere. The keep test needs only a compact summary of the population and continuum opacity in each species and frequency bin, together with the temperature dependence of excitation. Let $C_{jb}$ be the continuum threshold at depth $j$ and frequency bin $b$,
\begin{equation}
C_{jb}=\eta_{\rm sel}
\frac{\kappa_{\nu_b,{\rm c},j}+\sigma_{\nu_b,{\rm c},j}}
{1-\exp[-h\nu_b/(k_{\rm B}T_j)]}.
\end{equation}
\noindent Here $\eta_{\rm sel}=10^{-3}$ is the fixed selection fraction.
For ion stage $r$ of species $s$, the depth dependence is reduced to
\begin{equation}
Q_{srb} =
\max_j\left[
\frac{(n_{s,r}/U_{s,r})_j}
{\rho_j\,(\Delta\nu_{\rm D}/\nu)_{jsr}\,C_{jb}}
\right] .
\end{equation}
\noindent Up to fixed unit conversions, the central strength proxy for line $l$ is
\begin{equation}
R_l=C_0\,\frac{(gf)_l\,Q_{s(l)r(l)b(l)}}{\nu_{b(l)}} .
\end{equation}
\noindent Here $s(l)$, $r(l)$, and $b(l)$ are the species, ion stage, and frequency bin assigned to line $l$, and $C_0$ contains fixed line-opacity constants and unit conversions. The line is retained when both $R_l\geq 1$ and $R_l\exp[-E_l/(k_{\rm B}T_*)]\geq 1$, where $T_*$ is the deepest layer temperature. This conservative test avoids discarding a line that can matter in any layer.

The selection fraction is the one free number in this test, and the
converged structure is insensitive to it. Lowering $\eta_{\rm sel}$ by three
orders of magnitude moves the temperature over the 80 depth points by at most
5.9~K for the Sun and 1.8~K for a red giant, with the change essentially complete
by $10^{-4}$. Synthesizing from each converged atmosphere over 300 to 1000~nm,
the normalized flux changes by $2.1\times10^{-4}$ and $1.3\times10^{-4}$ rms,
below the residual against the original Kurucz calculation in
Section~\ref{sec:synthesis}. The runtime is insensitive to the threshold as well,
with the four solutions for each star within 4 percent of one another.

Compiled passes apply these tests once per atmosphere to the fixed source catalog and retain the selected records for subsequent iterations. The same pattern extends across the physical cycle. Independent work is compiled and parallelized, reusable inputs remain resident, and the physical sequence is preserved. Table~\ref{tab:atmosphere_speed} summarizes these execution changes.

The atmosphere offers less GPU parallelism than synthesis. Frequency-parallel blocks are interleaved with the equation of state, reductions, convection, correction, and remapping. These stages exchange a small depth state and must remain ordered, while the available wavelength batch is much smaller than in broad spectral synthesis. The production solver therefore uses multicore CPU kernels. Thread scaling saturates when memory traffic and the ordered stages dominate, with the saturation point set by the processor.

\subsection{Accuracy and performance}
\label{subsec:atmosphere_performance}

The first comparison in Figure~\ref{fig:synthesis_overview} holds the atmosphere fixed to isolate synthesis. In the second, each atmosphere solver produces its own structure at the same stellar labels before synthesis, testing whether atmosphere differences remain negligible in the final observable. For the hot dwarf, Sun, red giant, and K dwarf over 300--1000~nm, the 99th-percentile absolute normalized-flux differences are $(0.70,\,2.18,\,5.76,\,2.90)\times10^{-3}$. The combined atmosphere and synthesis calculation is thus in practical spectral parity with the original Kurucz programs across the tested regimes.

To isolate the cost of one physical pass from initialization and stopping, both implementations begin from the same structure and execute three iterations with identical physical inputs on the same AMD EPYC 9B45 CPU. Figure~\ref{fig:atmosphere_performance} summarizes the matched iteration walls, thread scaling, and red-giant stage breakdown. This design measures the execution of one physical pass rather than the choice of starting state or stopping policy.

For the hot dwarf, Sun, red giant, and K dwarf, the original Kurucz walls are 15, 13, 34, and 38~s per iteration. The corresponding 16-thread \paynezero\ walls are 2.1, 2.2, 4.7, and 5.3~s, giving factors of 6--7. The original program has no thread-parallel execution path, so its single-core measurement is its normal operating mode rather than an imposed restriction. That factor combines two
distinct effects, and the retained campaign separates them on the same host.
Against single-thread \paynezero, the original walls give an algorithmic factor
of 1.4 to 2.0, and moving \paynezero\ from one thread to sixteen gives a further
3.5 to 4.5. The algorithmic term is largest for the red giant, whose crowded
catalog makes the resident line data and the fused keep test worth most.

The two terms matter in different places. For one atmosphere the combined
factor is the applicable comparison, since the original program cannot use more than
one core and a converged atmosphere inside a fitting loop is a latency problem.
For bulk throughput the parallel term is available to both codes, because
independent stars can be distributed over cores without changing either
calculation. The per-star cost reduction in that regime is the algorithmic term
alone. Scaling in \paynezero\ saturates near 16 threads because memory traffic and ordered stages begin to dominate. In the measured configuration, the CPUs advance the ordered atmosphere calculation while the GPU handles the much broader synthesis parallelism.

The red-giant breakdown shows that the original wall is dominated by catalog input and line selection, while opacity is the largest remaining \paynezero\ cost. The gain therefore combines resident data, a parallel keep test, and compiled opacity kernels within one CPU--GPU workflow. Per-iteration acceleration leaves the number of iterations as the remaining multiplicative cost.

\section{Learned Atmosphere Initialization}\label{sec:atmosphere_emulator}

We reduce this remaining cost by learning the starting state while retaining the physical correction cycle. One cycle rebuilds pressure and populations, opacity, transfer, convection, and the remapped temperature and column-mass structure. It forms no global derivative matrix, so each iteration costs one physical pass. This economical update works well across the hot-dwarf, solar, red-giant, and K-dwarf tests. In our tests, more general root solvers improved robustness but required many physical passes.

Convergence still depends on whether repeated updates from the starting state approach the physical solution. We therefore judge an initializer by its physical restart rather than profile accuracy alone. Every comparison keeps the target labels and abundances fixed and changes only the starting atmosphere. The prediction is not accepted as an atmosphere. The physical solver must pass its convergence checks, while independent tests examine flux closure and the synthesized spectrum.

\subsection{Initialization as a fixed point problem}

For fixed requested values of $T_{\rm eff}$, $\log g$, $\xi$, and the abundances $\{[\mathrm{X}/\mathrm{H}]\}$, let $\mathbf{x}^{(n)}$ denote the complete numerical state needed to begin iteration $n$. Its physical projection contains the six profiles on the 80-layer production grid,
\begin{equation}
\mathbf{p}^{(n)}
=\left\{m_j,T_j,P_{{\rm gas},j},n_{e,j},
\kappa_{{\rm R},j},g_{{\rm rad},j}\right\}_{j=1}^{80}
\end{equation}
\noindent while the full state also carries the previous temperature correction, accumulated Rosseland-opacity information, and iteration controls. One complete Kurucz cycle acting on it is $\mathcal{G}$,
\begin{equation}
\mathbf{x}^{(n+1)}=\mathcal{G}\left(\mathbf{x}^{(n)}\right).
\end{equation}
 \noindent whose fixed point is the converged solver state,
\begin{equation}
\mathbf{x}_\star=\mathcal{G}\left(\mathbf{x}_\star\right).
\end{equation}
 \noindent Initialization means choosing $\mathbf{x}^{(0)}$. Near the fixed point the error $\mathbf{e}^{(n)}=\mathbf{x}^{(n)}-\mathbf{x}_\star$ evolves as
\begin{equation}
\mathbf{e}^{(n+1)}
\simeq
\mathbf{J}_{\mathcal{G}}(\mathbf{x}_\star)\mathbf{e}^{(n)} .
\end{equation}
\noindent Local convergence requires the spectral radius of $\mathbf{J}_{\mathcal{G}}$ to be below unity, which defines a local basin. Its boundary can be complicated, since temperature and pressure change the populations and opacity, opacity changes the next correction, remapping couples optical depth to column mass, and the convective boundary can move.

The fixed point residual
\begin{equation}
\mathcal{R}(\mathbf{x})
=\mathcal{G}(\mathbf{x})-\mathbf{x}
\end{equation}
\noindent measures whether the complete carried state is consistent under one pass. Close agreement between the six starting and converged profiles does not guarantee that the complete state changes little under the next pass. Small profile errors can still combine into a large pressure, opacity, or convection response. A useful initializer must therefore lie inside the basin of the physical iteration, not merely minimize profile error.

An Eddington grey atmosphere provides a generic starting structure, but it lacks the response to line blanketing, molecular equilibrium, and convection \citep{Mihalas1978,Gustafsson2008}. It can therefore lie outside the useful basin. A precomputed atmosphere grid supplies nearby converged starting models \citep{CastelliKurucz2003}, although a stored model is useful only when the physical iteration remains in the target solution's basin.

Training an initializer first requires a set of converged physical atmospheres. Faster physical iteration makes this initial set affordable. Alternating physical solutions and retraining then expands the corpus, with only converged solutions retained as training targets.

\subsection{A compact physical atmosphere representation}

\begin{table*}[t]
\centering
\caption{Learned atmosphere initializer families, sample sizes, input labels, and training envelopes. $N_{\rm corpus}$ is the number of independently converged atmospheres retained before splitting. The next column gives the training, validation, and held-out test counts. The public direct-abundance input contains 81 $[\mathrm{X}/\mathrm{H}]$ values, represented internally by $[\mathrm{Fe}/\mathrm{H}]$ and 80 non-iron $[\mathrm{X}/\mathrm{Fe}]$ coordinates.}
\label{tab:initializer_families}
\begingroup
\setlength{\tabcolsep}{5pt}
\renewcommand{\arraystretch}{1.20}
\compacttablefont
\begin{tabular}{@{}lllll@{}}
\toprule
\tcell{0.11\textwidth}{\textbf{Family}} &
\tcell{0.08\textwidth}{$N_{\rm corpus}$} &
\tcell{0.16\textwidth}{\textbf{Training / validation / test}} &
\tcell{0.27\textwidth}{\textbf{Input labels}} &
\tcell{0.30\textwidth}{\textbf{Training envelope}} \\
\midrule
\tcell{0.11\textwidth}{\textbf{Five-label}} &
\tcell{0.08\textwidth}{52,199} &
\tcell{0.16\textwidth}{50,000 / 2,000 / 199} &
\tcell{0.27\textwidth}{$T_{\rm eff}$, $\log g$, $[\mathrm{M}/\mathrm{H}]$,
$[\alpha/\mathrm{M}]$, $\xi$} &
\tcell{0.30\textwidth}{$4000\leq T_{\rm eff}\leq10500$ K,
$0.7\leq\log g\leq5.3$, $-2.5\leq[\mathrm{M}/\mathrm{H}]\leq0.5$,
$-0.1\leq[\alpha/\mathrm{M}]\leq0.5$,
$0.5\leq\xi\leq4.0$ km s$^{-1}$} \\
\tcell{0.11\textwidth}{\textbf{Eight-label CNO}} &
\tcell{0.08\textwidth}{53,824} &
\tcell{0.16\textwidth}{51,625 / 2,000 / 199} &
\tcell{0.27\textwidth}{$T_{\rm eff}$, $\log g$, $[\mathrm{M}/\mathrm{H}]$,
$[\alpha/\mathrm{M}]$, $\xi$, $[\mathrm{C}/\mathrm{M}]$,
$[\mathrm{N}/\mathrm{M}]$, $[\mathrm{O}/\mathrm{M}]$} &
\tcell{0.30\textwidth}{Five-label bounds, with $-0.5\leq[\mathrm{C}/\mathrm{M}],
[\mathrm{N}/\mathrm{M}],[\mathrm{O}/\mathrm{M}]\leq0.5$} \\
\tcell{0.11\textwidth}{\textbf{Direct abundance}} &
\tcell{0.08\textwidth}{82,016} &
\tcell{0.16\textwidth}{71,660 / 9,733 / 623} &
\tcell{0.27\textwidth}{$T_{\rm eff}$, $\log g$, $\xi$, and 81
$[\mathrm{X}/\mathrm{H}]$ abundances, represented by
$[\mathrm{Fe}/\mathrm{H}]$ and 80 non-iron
$[\mathrm{X}/\mathrm{Fe}]$ coordinates} &
\tcell{0.30\textwidth}{$4000\leq T_{\rm eff}\leq10500$ K,
$0.7\leq\log g\leq5.3$, $0.5\leq\xi\leq4.0$ km s$^{-1}$,
$-2.5\leq[\mathrm{Fe}/\mathrm{H}]\leq0.5$, and
$-0.5\leq[\mathrm{X}/\mathrm{Fe}]\leq0.5$} \\
\addlinespace[2pt]
\bottomrule
\end{tabular}
\endgroup
\end{table*}

The central design choice is to predict all six fields together. Their raw values span many orders of magnitude and obey positivity and monotonicity constraints. We therefore transform layer $j$ to
\begin{equation}
\begin{aligned}
\mathbf{u}_j=\bigg(&
\log_{10}\Delta m_j,\,
\log_{10}\frac{T_j}{T_{{\rm grey},j}},\,
\log_{10}P_{{\rm gas},j},\\
&\log_{10}n_{e,j},\,
\log_{10}\kappa_{{\rm R},j},\,
\operatorname{asinh}\frac{g_{{\rm rad},j}}{s_g}
\bigg),
\end{aligned}
\end{equation}
\noindent where $\Delta m_1=m_1$ and $\Delta m_j=m_j-m_{j-1}$ for $j>1$, and
\begin{equation}
T_{\rm grey}(\tau_{\rm R})=
T_{\rm eff}
\left[\frac{3}{4}\left(\tau_{\rm R}+\frac{2}{3}\right)\right]^{1/4}.
\end{equation}
\noindent Positive mass increments enforce monotonicity, logarithms preserve positivity, and the grey scaling removes most of the global temperature trend. The family-specific scale $s_g$ is the median absolute radiative acceleration in the training corpus. The signed transform retains both directions of radiative acceleration.
The transformed $80\times6$ atmosphere contains 480 coordinates. A joint principal component analysis, or PCA, reduces it to $K=160$ coefficients while retaining more than 99.999 percent of the standardized training variance.

A feedforward network maps the five- and eight-label inputs in Table~\ref{tab:initializer_families} to standardized PCA coefficients. For the direct-abundance family, an element-aware set encoder first combines each abundance with its element identity \citep{Zaheer2017}, then joins that summary to $T_{\rm eff}$, $\log g$, and $\xi$, which denote effective temperature, surface gravity, and microturbulence, before predicting the same coefficient representation. If $\widehat{\mathbf z}$ is the standardized network output, decoding reverses both coefficient and profile standardization,
\begin{equation}
\begin{aligned}
\widehat{\mathbf c}
&=\overline{\mathbf c}+\mathbf{s}_{c}\odot\widehat{\mathbf z},
\\
\operatorname{vec}(\widehat{\mathbf u})
&=\overline{\mathbf u}
+\mathbf{s}_{u}\odot\mathbf{B}_{K}^{\mathsf T}\widehat{\mathbf c},
\\
\mathbf{p}^{(0)}&=\mathcal{D}(\widehat{\mathbf{u}}).
\end{aligned}
\end{equation}
\noindent Here $\overline{\mathbf c}$ and $\mathbf{s}_c$ are the training mean and scale of the PCA coefficients. The corresponding profile quantities are $\overline{\mathbf u}$ and $\mathbf{s}_u$, while $\mathbf{B}_K$ is the retained $K$-component PCA basis. Hats identify network predictions. The operator $\operatorname{vec}$ flattens the depth-by-field profile array, $\odot$ denotes elementwise multiplication, and $\mathsf T$ denotes transpose. The decoder $\mathcal{D}$ reverses the physical transforms and returns the six starting profiles $\mathbf{p}^{(0)}$. The solver constructs the remaining carried state around these profiles. A joint basis preserves correlated variations across fields and depth, which matters because the physical solver responds to all six profiles together.

We evaluate the training loss after PCA decoding, so it constrains the full depth structure rather than coefficient errors alone. Four terms address distinct restart errors. $\mathcal{L}_{\rm prof}$ matches the values of all six decoded profiles. $\mathcal{L}_{\nabla}$ matches their changes between adjacent layers, so a profile at the correct level cannot keep the wrong depth dependence. $\mathcal{L}_{\tau}$ checks the mapping between column mass and Rosseland optical depth, while $\mathcal{L}_{\rm hse}$ checks hydrostatic pressure balance. The coefficients $\lambda$ set the relative weight of the last three terms,
\begin{equation}
\label{eq:initializer_loss}
\mathcal{L}
=\mathcal{L}_{\rm prof}
+\lambda_{\nabla}\mathcal{L}_{\nabla}
+\lambda_{\tau}\mathcal{L}_{\tau}
+\lambda_{\rm hse}\mathcal{L}_{\rm hse}.
\end{equation}
\noindent The profile and gradient terms use a robust smooth-L1 mismatch on the transformed fields. The optical-depth and hydrostatic terms are evaluated after reconstructing temperature, pressure, opacity, and column mass in physical units. They measure departures from
\begin{equation}
\frac{\mathrm{d}\tau_{\rm R}}{\mathrm{d}m}=\kappa_{\rm R},
\qquad
\frac{\mathrm{d}P_{\rm gas}}{\mathrm{d}m}\simeq g-g_{\rm rad}.
\end{equation}
\noindent These terms preserve the profile shape and two structural balances that strongly affect a physical restart. They do not replace radiative and convective equilibrium, which the physical solver must still establish. The five- and eight-label models use $\lambda_{\nabla}=0.1$ and $\lambda_{\tau}=\lambda_{\rm hse}=0.05$. The direct-abundance model uses $\lambda_{\nabla}=0.2$ with the same weights on the physical-consistency terms.

\subsection{Bootstrap, label coverage, and validation}
\label{sec:atmosphere_emulator_bootstrap}

We seeded a first label block with starting guesses from the Kurucz-A1 initializer \citep{Li2025}. Only atmospheres that converged with the physical solver entered the corpus. Those solutions trained a provisional initializer, whose predictions reduced the physical passes needed for the next block. Repeating physical solves and retraining produced the corpora summarized in Table~\ref{tab:initializer_families}.
The three released families are practical defaults rather than restrictions on the method. With each training atmosphere taking seconds to tens of seconds, a survey can generate tens of thousands in parallel for its own label range or abundance parameterization.

\begin{figure*}[t]
\centering
\includegraphics[width=\textwidth]{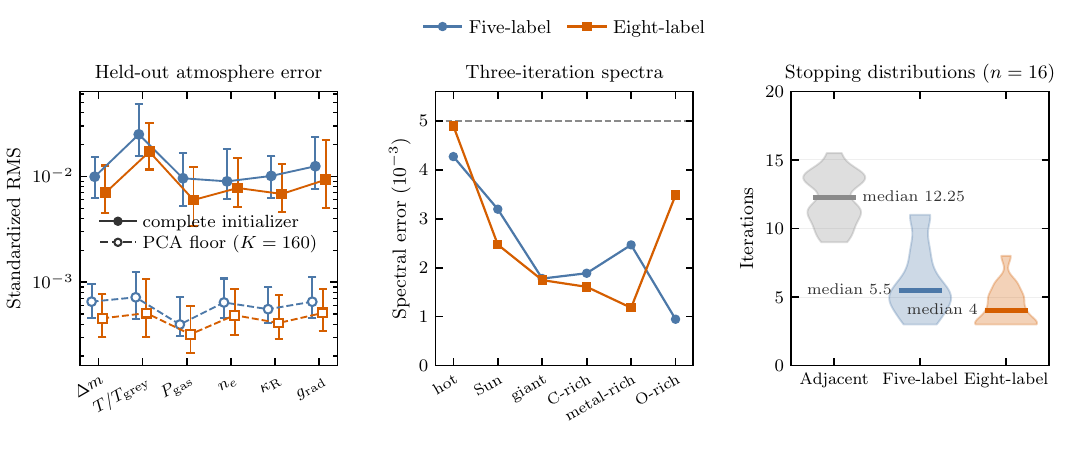}
\caption{Accuracy and iteration cost of the five- and eight-label atmosphere initializers. The left panel compares the $K=160$ PCA compression floor (dashed, open) with the complete held-out label-to-atmosphere error (solid, filled). Points are median standardized depth RMS values, with 16th--84th-percentile ranges. The middle panel gives a conservative spectral error bound after three physical-solver passes for six controls. By the triangle inequality, it sums the spectral change from passes three to five and the five-pass difference from the same stored reference. All lie below $5.1\times10^{-3}$. The right panel compares iteration counts from nearby stored-atmosphere starts with the two learned starts. Horizontal segments mark medians of about 12, 5.5, and 4 iterations, respectively.}
\label{fig:initializer_convergence}
\end{figure*}

The five-label family suits work in which a bulk metallicity and an alpha pattern describe the atmosphere adequately. Latin-hypercube marginals \citep{McKay1979} stratify every label range without an exponentially large tensor grid. A small imposed correlation between the $T_{\rm eff}$ and $\log g$ ranks samples common dwarf and giant structures more densely without restricting the corpus to an evolutionary track.

The eight-label family addresses abundance applications in evolved stars, where dredge-up changes surface C and N independently of the bulk metal and alpha patterns \citep{Salaris2015,Martig2016}. These changes redistribute atoms among CO, CN, and OH and can alter molecular opacity and atmospheric structure \citep{Tsuji1973,Gustafsson2008}. Conditional Latin-hypercube panels in C, N, and O are attached to sampled five-label parent states, and every resulting eight-label atmosphere is solved independently. In both lower-dimensional families $[\alpha/\mathrm{M}]$ adds to the bulk $[\mathrm{M}/\mathrm{H}]$ scaling for the alpha elements, except that the explicit $[\mathrm{O}/\mathrm{M}]$ coordinate replaces it for oxygen in the CNO family.
The direct-abundance family covers Li through Bi except Tc and Pm, together with Th and U, and has no bulk metallicity or alpha coordinate. Its internal change of coordinates does not alter the abundances passed to the physical solver.

We first validate the five- and eight-label families on six spectral controls spanning a hot dwarf, the Sun, a red giant, and C-enhanced, metal-rich, and O-enhanced mixtures. Individual abundance overrides are absolute $[\mathrm{X}/\mathrm{H}]$ values and replace the bulk scaling for that element.
The left panel of Figure~\ref{fig:initializer_convergence} separates two errors. Open symbols show the $K=160$ PCA compression floor, whose median standardized-coordinate RMS is $6.3\times10^{-4}$ for the five-label family and $4.9\times10^{-4}$ for the eight-label family. The complete prediction from labels through the network and PCA decoder gives $1.5\times10^{-2}$ and $1.1\times10^{-2}$, more than an order of magnitude above that floor. The learned label-to-atmosphere relation, rather than PCA compression, therefore sets the current accuracy.

The middle panel bounds the spectral error remaining after three solver passes by $|f_3-f_{\rm ref}|\leq |f_3-f_5|+|f_5-f_{\rm ref}|$. All six controls lie below $5.1\times10^{-3}$ for both released initializers. Three passes are a progress diagnostic rather than a stopping rule.

The right panel isolates the effect of the starting atmosphere in 16 matched held-out cases. To emulate a nearby node in a precomputed grid we selected converged structures near predeclared offsets of $\pm250$ K in $T_{\rm eff}$, $\pm0.25$ dex in $\log g$, and $\pm0.20$ dex in metallicity. Only the stored structure initializes the calculation, and every atmosphere is then solved at the exact target labels.

The median of the two grid-like starts is about 12 iterations, compared with 5.5 for the five-label initializer and 4 for the eight-label initializer. Combined with the six- to seven-fold per-iteration gain, the eight-label calculation is about 20 times faster in these matched cases. The retained original-program benchmark requests 30 iterations per batch. Four physical passes and the faster iteration give a gain of about 50 relative to that benchmark, while a nearby grid atmosphere gives the smaller and more realistic matched comparison above.

\begin{figure*}[t]
\centering
\includegraphics[width=\textwidth]{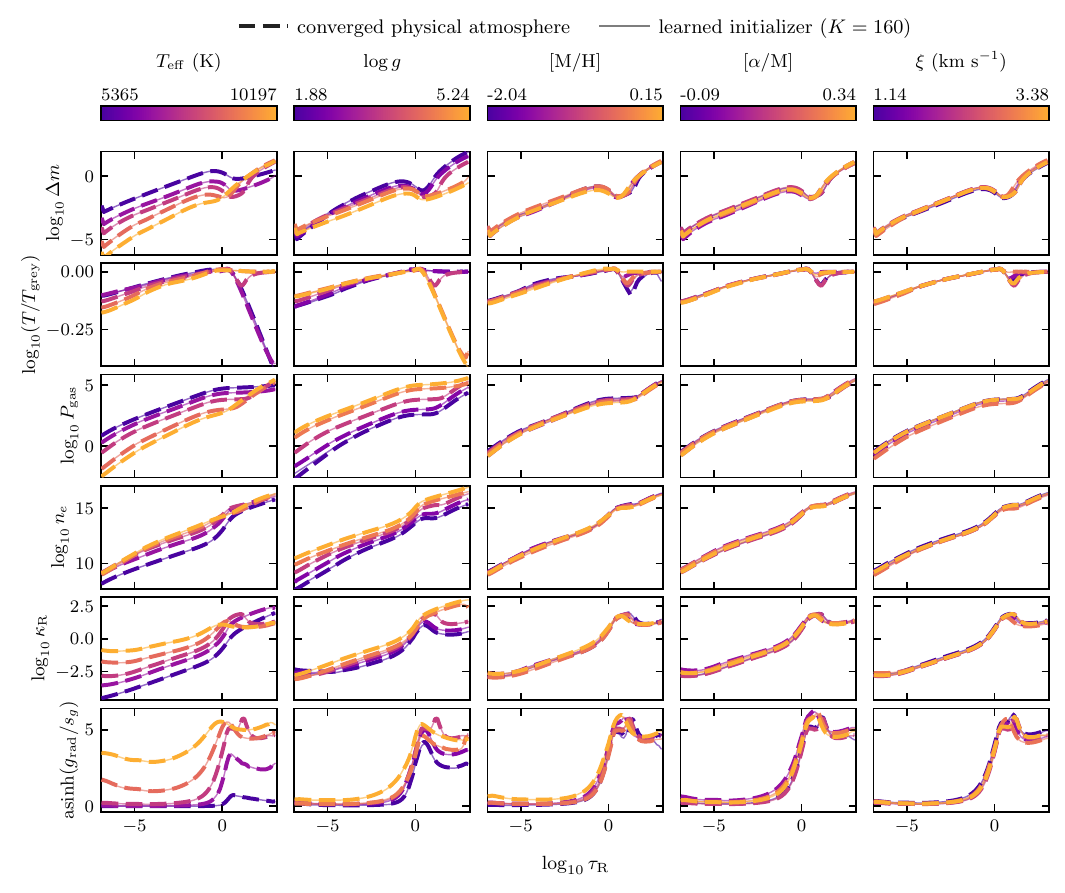}
\caption{Physical atmosphere structure and learned initialization. Columns vary $T_{\rm eff}$, $\log g$, $[\mathrm{M}/\mathrm{H}]$, $[\alpha/\mathrm{M}]$, or microturbulence as encoded by their color scales. Rows show six transformed depth profiles on a common Rosseland optical-depth grid. Dashed curves are converged physical atmospheres, and solid curves are the eight-label predictions used to initialize the physical solver.}
\label{fig:initializer_profiles}
\end{figure*}

Figure~\ref{fig:initializer_profiles} provides a direct profile-space check of the eight-label initializer. Solid predictions are compared with dashed converged physical profiles for 25 atmospheres whose C, N, and O values follow the corresponding bulk mixture. Each column varies one stellar label while the six rows trace the coupled temperature, pressure, density, opacity, and depth responses. This comparison tests the complete mapping from labels through the network and PCA decoder. Pooling the standardized residuals over all profiles, atmospheres, and depth points gives an RMS of 0.028 over the full depth range and 0.027 for $\log\tau_{\rm R}\geq1$.

For the direct-abundance family, the 623-object test split contains 313 internal and 310 external atmospheres. The internal atmospheres were excluded from network fitting and have 95th-percentile temperature and gas-pressure errors of 0.67 percent and 0.034 dex. A separate 16-case restart diagnostic converges in 8--14 physical passes, with a median near 9. In the application below, this network supplies the starting state and every retained atmosphere is then solved and validated physically.

This family presents a harder interpolation problem because three stellar parameters and 81 independently varying abundances define an 84-dimensional input. Its training corpus samples that space much more sparsely than the five- and eight-label corpora. Coordinates that group abundances by their chemical or opacity response may reduce the effective dimension in future initializers.

Each released initializer is supported only inside the training envelope of Table~\ref{tab:initializer_families}. Every family requires $T_{\rm eff}\geq4000$~K and holds a metallicity floor of $-2.5$~dex. Outside the envelope the released code warns and projects the initializer query onto it, leaving the labels passed to the physical solver unchanged, so the convergence test still decides acceptance.
These limits apply to the initializers alone. Synthesis and the
physical atmosphere solver are bounded instead by the supported Kurucz physics,
which is one-dimensional and LTE.

\section{Normalized Spectrum Fitting}\label{sec:fitter}

If atmosphere and synthesis calculations are fast enough, the physical forward model itself can be evaluated inside a spectrum optimizer. Direct synthesis can explore labels repeatedly, while converging a new atmosphere at every trial would dominate the calculation. We therefore use the initializer to supply the starting depth structure, recompute populations, line opacity, and radiative transfer at every trial, and reserve atmosphere convergence for the candidate minimum and any material correction. No label-to-flux spectral emulator enters the fit.
We first test this strategy on a controlled normalized spectrum whose generating and fitted models are identical. The test isolates label recovery and atmosphere correction before introducing line-data and survey-instrument mismatch.

\begin{figure*}[t!]
\centering
\includegraphics[width=\textwidth]{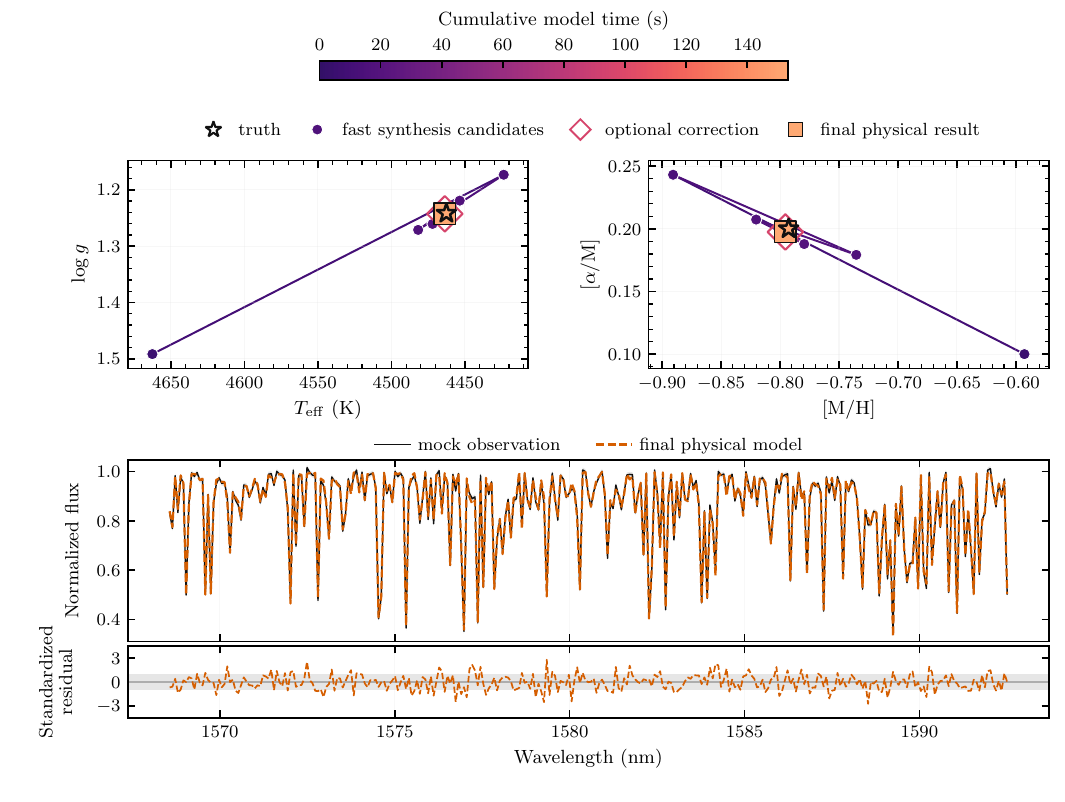}
\caption{Controlled five-label fit to a normalized mock spectrum with ${\rm S/N}=100$ and known truth. The upper panels trace accepted fast-search candidates (circles), the optional correction informed by the first converged atmosphere (diamond), and the final solution from a converged atmosphere and synthesis (square) in the $T_{\rm eff}$--$\log g$ and $[\mathrm{M}/\mathrm{H}]$--$[\alpha/\mathrm{M}]$ planes. Color gives cumulative model time, and the hollow star marks the truth. The lower panels compare the mock observation with the final physical model over a representative wavelength interval, with shaded $1\sigma$ noise bands. The recovered spectrum and labels use direct synthesis without a label-to-flux emulator.}
\label{fig:fitter_convergence}
\end{figure*}

The fast search begins at the initial label vector. One direct spectrum and one bounded perturbation per label define a local spectral response. A weighted linear least-squares step proposes a bounded label update, and a direct-synthesis line search retains it only when the objective improves \citep{NocedalWright2006}. The released fitter rebuilds this response at each iteration. The short Figure~\ref{fig:fitter_convergence} diagnostic reuses its initial response between successive accepted steps. We obtain the response by finite differences because the complete atmosphere-to-spectrum calculation does not yet form one PyTorch computational graph. An analytic response could replace these differences without changing the physical atmosphere calculation. Here $N_{\rm good}$ is the number of retained pixels. The objective $Q$ is the mean chi-square per valid pixel. The fast loop stops when no bounded line-search trial lowers $Q$.

The first converged atmosphere does more than test this candidate. Let $f_{\rm fast}$ and $f_{\rm phys}$ denote the spectra before and after physical atmosphere convergence at the fast minimum. We hold $\Delta f=f_{\rm phys}-f_{\rm fast}$ fixed over one bounded local step and rebuild the spectral response for every active label. A Gauss--Newton step then solves the weighted linear least-squares problem for the combination of label changes predicted to reduce the pixel residuals. Every active label participates in this correction. Fits that expose element-by-element abundance coordinates may therefore adjust them after the first physical calculation.

\begin{table}[t!]
\centering
\begin{minipage}{\columnwidth}
\centering
\textbf{Algorithm 1. Controlled direct spectrum fit}\par
\vspace{2pt}
\noindent\rule{\columnwidth}{0.5pt}
\vspace{2pt}
\begin{algorithmic}[1]
\Require flux $y_p$, weights $w_p$, initial labels $\boldsymbol{\theta}_0$,
threshold $\delta Q_{\min}$
\Ensure labels $\boldsymbol{\theta}_\star$, atmosphere state $\mathbf{x}_\star$,
spectrum $f_\star$
\State Define $Q(f)\gets N_{\rm good}^{-1}\sum_{p\in{\rm good}}w_p(y_p-f_p)^2$
\State $\boldsymbol{\theta}\gets\boldsymbol{\theta}_0$
\Repeat
 \State Evaluate direct $f_{\rm fast}$ and its local response
 \State Propose a bounded least-squares step
 \State Retain it only if direct synthesis lowers $Q$
\Until{no bounded trial lowers $Q$}
\State $\boldsymbol{\theta}_{\rm f}\gets\boldsymbol{\theta}$
\State Solve $\mathbf{x}_{\rm phys}$ at $\boldsymbol{\theta}_{\rm f}$ and synthesize $f_{\rm phys}$
\State Set $(\boldsymbol{\theta}_\star,\mathbf{x}_\star,f_\star)
 \gets(\boldsymbol{\theta}_{\rm f},\mathbf{x}_{\rm phys},f_{\rm phys})$
\State Fix $\Delta f\gets f_{\rm phys}-f_{\rm fast}$
\State Rebuild the response
\State Propose bounded correction $\boldsymbol{\theta}_{\rm p}$
\If{its predicted decrease exceeds $\delta Q_{\min}$}
 \State Solve $\mathbf{x}_{\rm prop}$ at $\boldsymbol{\theta}_{\rm p}$ and synthesize $f_{\rm prop}$
 \If{$Q(f_{\rm prop})<Q(f_{\rm phys})$}
 \State Set $(\boldsymbol{\theta}_\star,\mathbf{x}_\star,f_\star)
 \gets(\boldsymbol{\theta}_{\rm p},\mathbf{x}_{\rm prop},f_{\rm prop})$
 \EndIf
\EndIf
\State \Return $(\boldsymbol{\theta}_\star,\mathbf{x}_\star,f_\star)$
\end{algorithmic}
\vspace{2pt}
\noindent\rule{\columnwidth}{0.5pt}
\end{minipage}
\end{table}

This local correction assumes only that $\Delta f$ varies slowly across the local trust region. We send the proposal to a second converged atmosphere only when its predicted decrease in $Q$ exceeds $\delta Q_{\min}$. For this controlled diagnostic, $\delta Q_{\min}=0.005$ in mean chi-square per valid pixel. We accept the proposal only when that physical spectrum lowers the same objective. Algorithm~1 summarizes the single-correction path used for this controlled diagnostic. The released refinement repeats this calculation only when the predicted decrease in $Q$ exceeds $\delta Q_{\min}$ and can continue bounded corrections up to its cap on physical evaluations.

The controlled example is a held-out non-solar K giant with offsets in all five initial labels and independent Gaussian noise at ${\rm S/N}=100$. It uses a normalized 1500--1700~nm spectrum on a logarithmic grid with $R_{\rm grid}=20{,}000$, with no instrumental line-spread function or other broadening. The fit minimizes the mean chi-square per valid pixel and omits the masks, continuum, velocity, broadening, and varying inverse variances needed for observed spectra. Its noiseless truth comes from a converged atmosphere and direct synthesis, so both label recovery and the fast-stage approximation are known. The final physical model recovers $T_{\rm eff}$ within 2~K, $\log g$ and both abundance coordinates within 0.003~dex, and microturbulence within 0.04~km~s$^{-1}$.

These are the precisions of a fit whose forward model generated the data, so they measure recovery rather than accuracy. To separate the two we fit the original Kurucz spectra of the four representative stars in Figure~\ref{fig:synthesis_overview} with the same model at the same signal-to-noise, which puts a residual the model cannot reproduce into the data. Across the hot dwarf, the Sun, the red giant, and the K dwarf the induced bias reaches 4.6~K in effective temperature, 0.013~dex in surface gravity, and 0.011~dex in metallicity. It is largest for the red giant and falls when the continuum is given more freedom. The difference from the original programs therefore corresponds to a few kelvin, above the recovery precision quoted here, and bounds agreement with that implementation.

Figure~\ref{fig:fitter_convergence} traces the accepted label trajectory in the upper panels and compares the final physical spectrum with the mock observation in the lower panels, where the residuals match the injected noise. The first converged atmosphere is already close to the mock truth. To demonstrate the correction path, this diagnostic nevertheless takes one bounded label step and confirms it with a second converged atmosphere. In routine use, the refinement takes such a step only when its predicted decrease in $Q$ exceeds $\delta Q_{\min}$. The five-label search takes about 20~s on one H100, while the converged-atmosphere checks run on multicore CPUs and dominate the remaining wall time. This near-exact recovery establishes the optimizer on a known physical model. We next address the model--data mismatch present in observed spectra through line-list calibration.

\section{Line List Calibration}\label{sec:linecal}

The controlled mock establishes recovery when the forward model is internally consistent. Observed spectra also test the physical inputs to that model. Following established near-infrared astrophysical line calibration, we ask whether the differentiable calculation can update oscillator strength and the three damping families jointly \citep{MelendezBarbuy1999,Shetrone2015,Smith2021}. Each proposed correction is constrained to change the spectrum through line opacity and radiative transfer. We evaluate the corrections against high-resolution Fourier transform spectrometer (FTS) atlases of the Sun and Arcturus.

For line component $l$ assigned to correction group $q(l)$, we write the calibrated quantities as
\begin{equation}
\begin{aligned}
(gf)'_l &= (gf)_l 10^{\delta_{q(l),gf}},\\
\gamma'_{l,\beta} &= \gamma_{l,\beta}10^{\delta_{q(l),\beta}},
\qquad \beta\in\{\mathrm{vdW},\mathrm{rad},\mathrm{S}\}.
\end{aligned}
\label{eq:atomic_calibration_parameters}
\end{equation}
\noindent Here $\gamma_{l,\beta}$ is the original damping coefficient for line component $l$ and family $\beta$. The three families are van der Waals, radiative, and Stark damping, while the prime denotes the calibrated value. The fourth fitted atomic-parameter type is $\log(gf)$. At fixed population, oscillator strength scales the frequency-integrated opacity, while damping redistributes opacity from the core into the wings \citep{Barklem2000,Gray2005}. Their responses overlap in saturated and blended features.

PyTorch records the path from each atomic correction through opacity, radiative transfer, broadening, sampling, continuum profiling, and the final objective. A reverse pass gives derivatives for every active correction together, without finite differences \citep{Baydin2015}. Independent line-by-line radiative-transfer work demonstrates the same computational principle in spectral inference \citep{Kawahara2022}. The fixed atmosphere, continuous and molecular opacity, and catalog geometry remain resident on the GPU.

Because line centers enter opacity deposition through integer grid placement, the graph does not yet provide a valid wavelength derivative. Wavelengths are therefore fixed. Lower excitation energies also remain outside this four-parameter proof of concept, which calibrates the quantities in Equation~\ref{eq:atomic_calibration_parameters}.

We use the 1500--1700 nm interval to connect directly to the APOGEE wavelength range. We perform the calibration against these FTS atlases rather than APOGEE pixels. The synthesis calculation uses $R_{\rm grid}=300{,}000$ before sampling onto each atlas grid. We use the Livingston and Wallace solar center-of-disk atlas and two Hinkle et al. Arcturus epochs \citep{LivingstonWallace1991,Hinkle1995,Smith2021}. For the solar calculation we adopt $T_{\rm eff}=5777$ K, $\log g=4.44$, and $\xi=0.7$ km s$^{-1}$. For Arcturus we adopt $T_{\rm eff}=4286$ K, $\log g=1.66$, $[\mathrm{M}/\mathrm{H}]=-0.52$, $[\alpha/\mathrm{M}]=0.30$, and $\xi=1.7$ km s$^{-1}$ \citep{Ramirez2011,Smith2021}.
We adopt the element-by-element Arcturus mixture tabulated by \citet{Smith2021}. That compilation draws most values from \citet{Ramirez2011} and supplements them from other element-specific studies.

Each standard's converged physical atmosphere remains fixed during the inner calibration. We adopt a Gaussian dispersion of 1.55~km~s$^{-1}$ for the Sun. For Arcturus we adopt a Gaussian dispersion of 2.5~km~s$^{-1}$ and $v\sin i=2.0$~km~s$^{-1}$. The supplied Doppler factors and a residual registration correction place the atlases in the model rest frame.

\begin{table*}[t!]
\noindent
\begin{minipage}[t]{\dimexpr(\textwidth-\columnsep)/2\relax}
\centering
\captionof{table}{Reduction relative to the original line data in the telluric-downweighted mean squared atlas--model flux difference. The first five rows fit each standard independently. The final row fits the Sun and Arcturus with one common set of line-parameter corrections.}
\label{tab:linecal_ablation}
\begingroup
\setlength{\tabcolsep}{3.5pt}
\renewcommand{\arraystretch}{1.12}
\compacttablefont
\par\centering
\hspace*{-0.10\linewidth}%
\begin{tabular}{@{}lrr@{}}
\toprule
\textbf{Calibration case} & \textbf{Sun} & \textbf{Arcturus} \\
\midrule
$\log(gf)$ & 78.2\% & 56.8\% \\
van der Waals damping & 36.9\% & 7.7\% \\
radiative damping & 19.0\% & 5.4\% \\
Stark damping & 26.6\% & 2.7\% \\
all four types & 90.1\% & 64.9\% \\
\midrule
shared Sun--Arcturus fit & 83.5\% & 56.0\% \\
\addlinespace[2pt]
\bottomrule
\end{tabular}
\endgroup
\end{minipage}
\hspace{\columnsep}%
\begin{minipage}[t]{\dimexpr(\textwidth-\columnsep)/2\relax}
\centering
\captionof{table}{Contents and application of the released common Sun--Arcturus line-list calibration. Source identities bind the corrections to the intended catalog. Each linked transition group stores four logarithmic corrections, from which the calibrated component values are obtained.}
\label{tab:linecal_overlay}
\begingroup
\setlength{\tabcolsep}{3pt}
\renewcommand{\arraystretch}{1.12}
\compacttablefont
\begin{tabular}{@{}lll@{}}
\toprule
\tcell{0.24\linewidth}{\textbf{Quantity}} &
\tcell{0.38\linewidth}{\textbf{Application}} &
\tcell{0.27\linewidth}{\textbf{Physical role}} \\
\midrule
\tcell{0.24\linewidth}{source identity} & \tcell{0.38\linewidth}{catalog digest and\\component map} & \tcell{0.27\linewidth}{selects the exact\\source transitions} \\
\tcell{0.24\linewidth}{linked group} & \tcell{0.38\linewidth}{component $\rightarrow$\\fitted group} & \tcell{0.27\linewidth}{shares one correction\\among linked lines} \\
\midrule
\tcell{0.24\linewidth}{$\log(gf)$} & \tcell{0.38\linewidth}{$\log(gf)_{\rm cal}=$\\$\log(gf)_{\rm base}+\delta_{q,gf}$} & \tcell{0.27\linewidth}{integrated line\\opacity} \\
\tcell{0.24\linewidth}{van der Waals\\damping} & \tcell{0.38\linewidth}{$\gamma_{\rm cal}=$\\$\gamma_{\rm base}10^{\delta_{q,\rm vdW}}$} & \tcell{0.27\linewidth}{neutral-collision\\wings} \\
\tcell{0.24\linewidth}{radiative damping} & \tcell{0.38\linewidth}{$\gamma_{\rm cal}=$\\$\gamma_{\rm base}10^{\delta_{q,\rm rad}}$} & \tcell{0.27\linewidth}{natural line\\width} \\
\tcell{0.24\linewidth}{Stark damping} & \tcell{0.38\linewidth}{$\gamma_{\rm cal}=$\\$\gamma_{\rm base}10^{\delta_{q,\rm S}}$} & \tcell{0.27\linewidth}{electron-pressure\\wings} \\
\bottomrule
\end{tabular}
\endgroup
\end{minipage}
\end{table*}

At every evaluation we profile a smooth multiplicative continuum using pixels near the continuum in both the atlas and the baseline model. Tight coefficient bounds prevent this adjustment from absorbing line-profile mismatch. Companion telluric-transmission arrays define relative pixel weights and mask unreliable wavelengths.

For each atlas, we minimize the mean squared flux difference after telluric downweighting and continuum profiling. The independent Arcturus solution fits the summer and winter epochs with equal weight, while the independent solar solution uses only the solar atlas. These fits measure the calibration capacity of each standard. We then fit one shared Sun--Arcturus solution, giving equal total weight to each star and equal weight to the two Arcturus epochs. We use this common calibration for APOGEE.

We include neutral and singly ionized atomic records in the interval for every species represented in the adopted abundance vector. We first preserve a small set of curated anchor complexes. A strength-ordered greedy pass then groups each remaining component with unused lines of the same species and ion stage within 2.0~cm$^{-1}$ in lower excitation and 0.035~nm in wavelength. The 42,886 catalog records form 25,281 groups and give 101,124 fitted scalar corrections. All components in one group share its four corrections.

\begin{figure*}[!t]
\centering
\includegraphics[width=\textwidth]{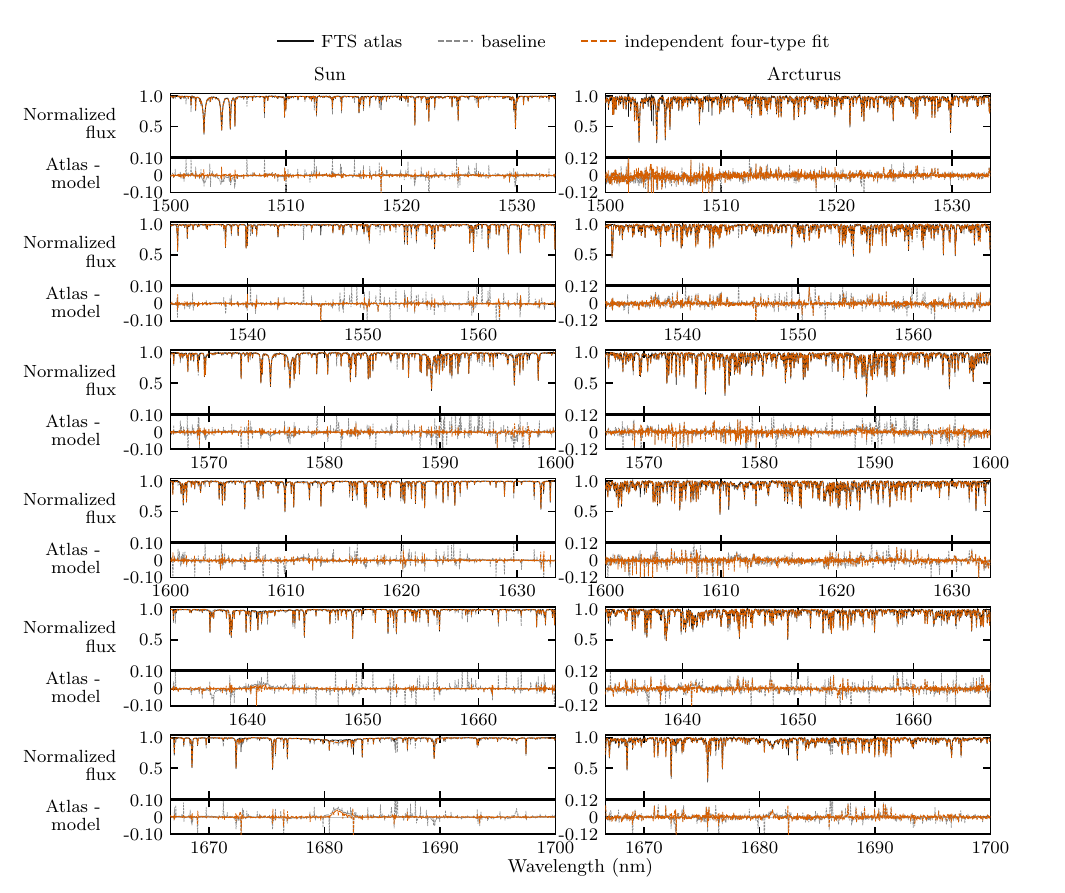}
\caption{Full-band calibration of the solar and Arcturus FTS atlases in six contiguous wavelength strips. For each strip, the upper panel shows normalized flux and the lower panel shows atlas minus model. Black is the atlas, grey dashed uses the original line data, and orange dashed uses the independent four-parameter line fit for that standard. Residual panels share a fixed range within each star. The Arcturus column shows the winter atlas, while its fit weights the summer and winter epochs equally. The calibration contracts the residual forest across the full interval for both standards.}
\label{fig:linecal_overview}
\end{figure*}

We optimize the complete 200 nm interval jointly with PyTorch L-BFGS \citep{NocedalWright2006}. This gradient-based method uses recent gradients to estimate curvature, while a strong-Wolfe line search accepts steps that lower the objective with a consistent local slope. Smooth parameter transforms enforce the physical bounds. Joint optimization lets blended pixels constrain several transitions and avoids splitting a line profile among independently fitted wavelength windows.
With 101,124 active corrections, the reverse pass avoids synthesizing one perturbed spectrum per parameter. Starting from the key-aligned mean of the independent solutions, 40 shared Sun--Arcturus updates take 64~s on one H100.

Table~\ref{tab:linecal_ablation} compares the single-parameter controls with each standard's independent four-parameter fit. The reductions are not additive because the parameter types can explain part of the same residual. Oscillator strength carries most of the gain in both stars, while the four-parameter fit improves further in each case. Damping adds more leverage in the solar spectrum, where the higher photospheric pressure produces stronger pressure-broadened wings \citep{Barklem2000,Gray2005}. In Arcturus, the fit empirically assigns most of the gain to line strength. Its crowded molecular spectrum provides fewer isolated wing constraints.

Figure~\ref{fig:linecal_overview} shows that the objective reduction is not confined to a few isolated features. The residual forest contracts across the displayed interval for both atlases. The change is stronger for the Sun, while Arcturus retains larger structured residuals in its more crowded spectrum.

\begin{figure*}[t!]
\centering
\begin{minipage}{\textwidth}
\centering
\includegraphics[width=0.90\textwidth,trim=0 36bp 0 8bp,clip]{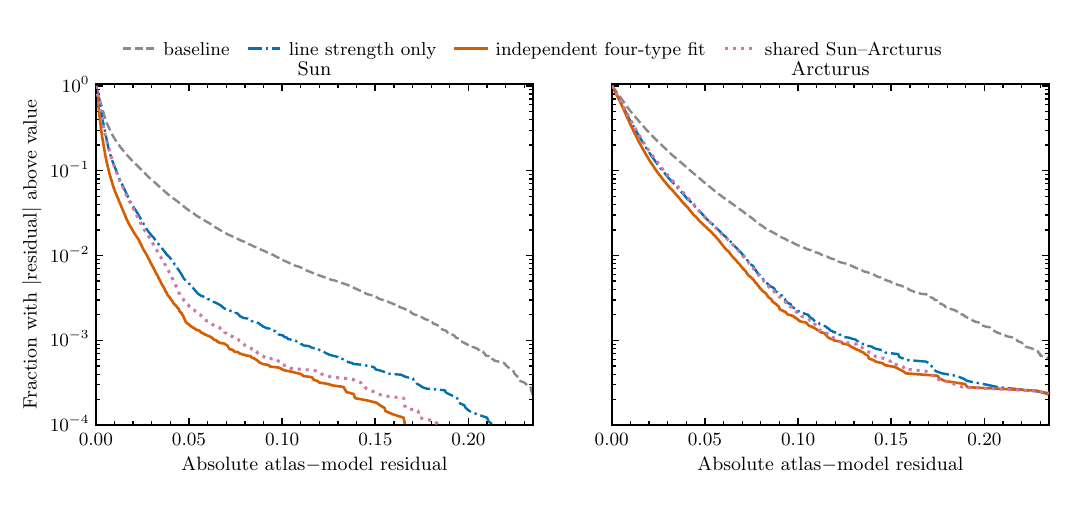}
\captionof{figure}{Absolute normalized-flux residual distributions for the Sun and Arcturus. At each horizontal value, the vertical axis is the fraction of retained atlas pixels with a larger absolute atlas--model residual. Grey dashed uses the original line data, blue dash-dotted fits oscillator strength alone, orange solid fits all four line parameters independently for each standard, and magenta dotted uses the shared Sun--Arcturus fit. All calibrations contract the residual tail, with the four-parameter fits giving the largest reduction. Fractions are unweighted over pixels retained after telluric masking.}
\label{fig:linecal_residual_distribution}
\end{minipage}
\vspace{5pt}
\begin{minipage}{\textwidth}
\centering
\includegraphics[width=0.90\textwidth]{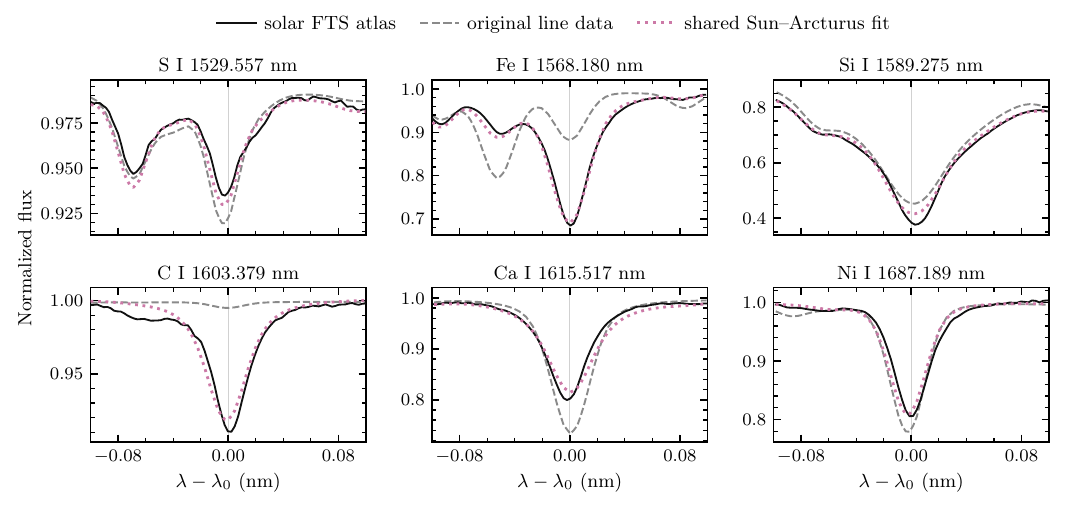}
\captionof{figure}{Six solar center-of-disk FTS line profiles evaluated with the retained shared Sun--Arcturus calibration. Wavelength is measured from the labeled transition's catalog laboratory center. Black is the atlas, grey dashed uses the original line data, and magenta dotted uses the shared fit of oscillator strength and three damping terms. The examples span six atomic species and show improvements in both line depth and profile shape.}
\label{fig:linecal_profiles}
\end{minipage}
\end{figure*}

Figure~\ref{fig:linecal_residual_distribution} shows unweighted residual distributions for the solar atlas and the displayed Arcturus winter epoch. Table~\ref{tab:linecal_ablation} instead reports telluric-weighted mean-square reductions over the calibration data, including both Arcturus epochs. Both the independent and shared fits contract the long residual tail.

Figure~\ref{fig:linecal_profiles} zooms in on six individual line profiles. The solar examples are selected algorithmically from the shared Sun--Arcturus fit and are not refitted locally. They are attributable, wavelength-separated transitions from different species. The shared calibration improves both line depths and profile shapes across these examples.

The fit has $10^5$ free scalars, so we ask whether the data constrain them.
The three atlas spectra retain 159,580 pixels after telluric masking, so
the nominal ratio is 0.63 fitted scalars per retained pixel. A line too weak to
change the spectrum contributes no gradient, and its correction stays at its
initial value. In the retained shared solution, 8,502 of the 101,124 scalars
depart from the base catalog by more than 0.01 dex and 5,396 by more than 0.05
dex. The effective ratio is therefore 0.05 constrained
corrections per retained pixel.
Beyond the bounds imposed by smooth parameter
transforms, the grouping of 42,886 components into 25,281 shared corrections is
the structural regularizer, and no explicit penalty is applied. The sparsity of
the solution follows from the data rather than from a prior.

We also evaluate the corrections against an independent observation.
Arcturus was recorded in two epochs. Fitting each parameter family on
one epoch alone and scoring it on the other, with the production bounds and
stopping tolerances, the four-type advantage over the
oscillator-strength-only control is 7--8 percentage points on the fitted
epoch at matched budgets, the size of the Table~\ref{tab:linecal_ablation}
gap. On the held-out epoch the advantage is about 2--5 points.
The test holds the
star fixed, so it bounds the absorption of observational noise rather than
of model error. Errors in the atmosphere, the continuum treatment, and missing
opacity are common to both epochs and transfer between them. Separating those
requires standards spanning different stellar
parameters and abundances, which we discuss in
Section~\ref{sec:discussion}.
These are model-dependent standard-star calibrations, not laboratory measurements \citep{Heiter2021}, and each standard-star atmosphere remains fixed during the inner fit.

We distribute the shared Sun--Arcturus calibration as a compact correction file. Applying it to the matching base catalog produces the calibrated line parameters summarized in Table~\ref{tab:linecal_overlay}. The \texttt{write\_substituted\_catalog} helper can write those values as a complete replacement catalog, while the default catalog remains unchanged. Independent solar and Arcturus fits remain available as diagnostic products.

The APOGEE line list carries astrophysical oscillator strengths adjusted
against FTS spectra of the same two standards, so it provides an external
comparison. Data Release 17 released two versions, tuned separately with a
Synspec and with a Turbospectrum synthesis from the same starting atomic data
and the same two stars, following the procedure of \citet{Smith2021}
\citep{Abdurrouf2022}. Matching on species and vacuum-converted wavelength gives
17,178 transitions whose starting oscillator strength is identical to ours. Our
corrections correlate with the two released versions at $+0.44$ and $+0.56$,
while those two versions correlate with each other at $+0.71$.
Appendix~\ref{app:linelist_comparison} gives the comparison and the differences
between the analyses.

A production calibration should share information across several standards \citep{Heiter2015,Jofre2015,Smith2021}. After a substantial opacity update, it should also repeat the atmosphere and line-list solutions.
Some large residuals remain because the current parameterization cannot move line centers or create missing transitions. We explored a residual-guided subset of predicted atomic transitions, but held-out tests did not justify promotion. Predicted transitions therefore require validation across independent stars before inclusion in a general line list.

\section{Application to APOGEE}\label{sec:apogee}

\begin{figure*}[t]
\centering
\includegraphics[width=\textwidth]{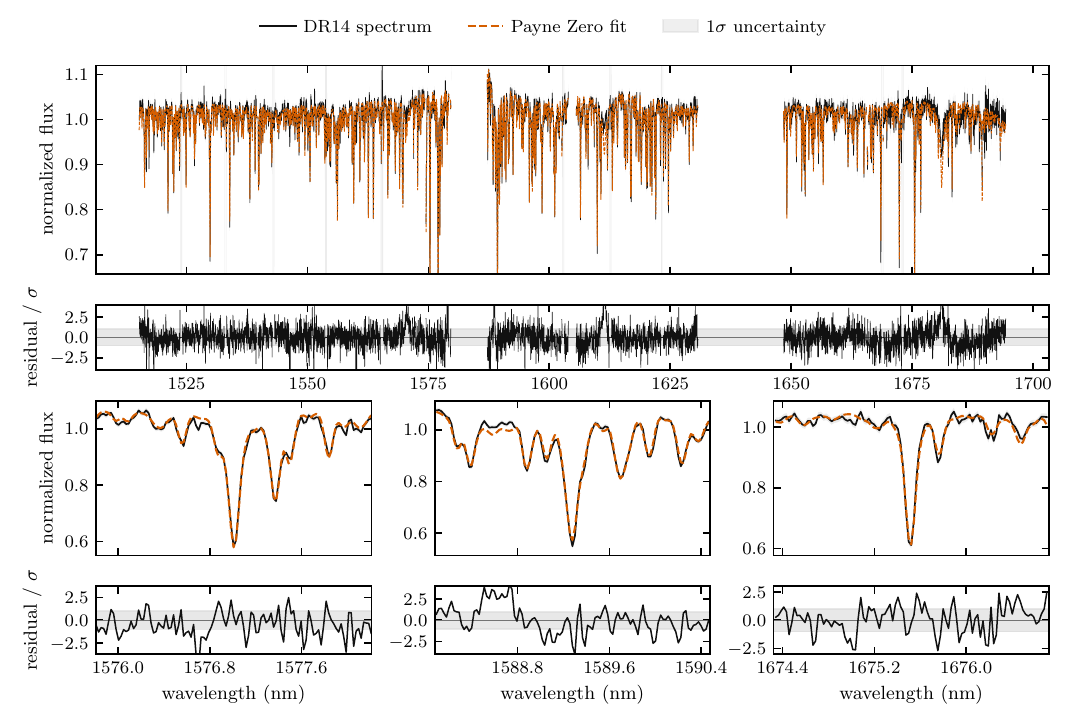}
\caption{Representative APOGEE DR14 fit after the converged atmosphere calculation. Black is the observed normalized spectrum, orange dashed is the profiled \paynezero\ model after velocity shift, broadening, line-spread function convolution, and pixel sampling, and grey bands show the adopted $1\sigma$ uncertainty. The upper spectrum and residual panels cover all three detectors. The three lower pairs expand one line-rich interval per detector. Residuals are shown in units of the pixel uncertainty.}
\label{fig:apogee_spectrum}
\end{figure*}

We now ask whether the same calculation can fit reduced survey spectra directly. We combine physical fitting with the retained standard-star line calibration and apply it to APOGEE Data Release 14 (DR14). The fiber-fed near-infrared spectrograph and its delivered resolution are described by \citet{Wilson2019}. We use this release because it provides the combined spectra, instrument calibrations, and giant-star analysis products needed for the controlled comparison \citep{Abolfathi2018,Holtzman2018}. It also publishes photometrically calibrated effective temperatures and asteroseismically calibrated giant surface gravities, which the fit below retains so that the chemical and instrumental terms are isolated. Labels from the APOGEE Stellar Parameters and Chemical Abundances Pipeline (ASPCAP) provide the external comparison \citep{GarciaPerez2016,Jonsson2018}.

For the survey demonstration, DR14 calibrates its raw spectroscopic $T_{\rm eff}$ to a low-reddening photometric scale and giant $\log g$ to asteroseismic gravities and fits microturbulence as a separate giant-star coordinate \citep{Holtzman2018}. We do not derive a new relation between spectroscopic and photometric temperature. For this proof of concept, we retain the published $T_{\rm eff}$, $\log g$, and $\xi$ to isolate the many-element and instrument problem. A production analysis could instead fit the spectrum jointly with a broadband spectral energy distribution and parallax \citep{Cargile2020}. Appendix~\ref{app:stellar_label_sensitivity} repeats the synthesis search with $T_{\rm eff}$, $\log g$, and $\xi$ free to test the sensitivity of the giant-branch, red-clump, and abundance morphology to this choice.

We fit $[\mathrm{Fe}/\mathrm{H}]$, $[\mathrm{X}/\mathrm{Fe}]$ for C, N, O, Mg, Al, Si, S, Ca, Ti, Mn, and Ni, residual velocity $v_r$, and effective Gaussian broadening $v_b$. For every retained element, $[\mathrm{X}/\mathrm{H}]=[\mathrm{Fe}/\mathrm{H}]+[\mathrm{X}/\mathrm{Fe}]$, while unselected metals follow $[\mathrm{Fe}/\mathrm{H}]$.
This frees 12 of the 81 abundance coordinates that the forward model
accepts. The remaining 69 follow the
fitted $[\mathrm{Fe}/\mathrm{H}]$, so every element still enters the equation of
state, the electron supply, molecular equilibrium, and line blanketing at the
requested mixture, and the atmosphere is converged for that full composition. We
free the 12 coordinates whose features are strongest in this interval. Further
elements are accessible in the H band and could be added, since each additional
coordinate costs one further response spectrum rather than a larger grid.

Synthesis returns native-grid total and continuum fluxes. We map both to the observed pixels before normalization,
\begin{equation}
f_p(\boldsymbol{\psi})=
\frac{\left\{\mathcal{K}\mathcal{S}(v_r)\mathcal{B}(v_b)
F_\lambda(\boldsymbol{\theta})\right\}_p}
{\left\{\mathcal{K}\mathcal{S}(v_r)\mathcal{B}(v_b)
F_{\lambda,{\rm c}}(\boldsymbol{\theta})\right\}_p},
\qquad
\boldsymbol{\psi}=(\boldsymbol{\theta},v_r,v_b).
\label{eq:apogee_forward}
\end{equation}
\noindent where $\mathcal{B}$ acts on the logarithmic synthesis grid, $\mathcal{S}$ applies the Doppler shift, and $\mathcal{K}$ applies the wavelength-dependent line-spread function and native APOGEE sampling. Convolving both before division preserves the nonlinear normalization.

\begin{figure*}[t]
\centering
\includegraphics[width=\textwidth]{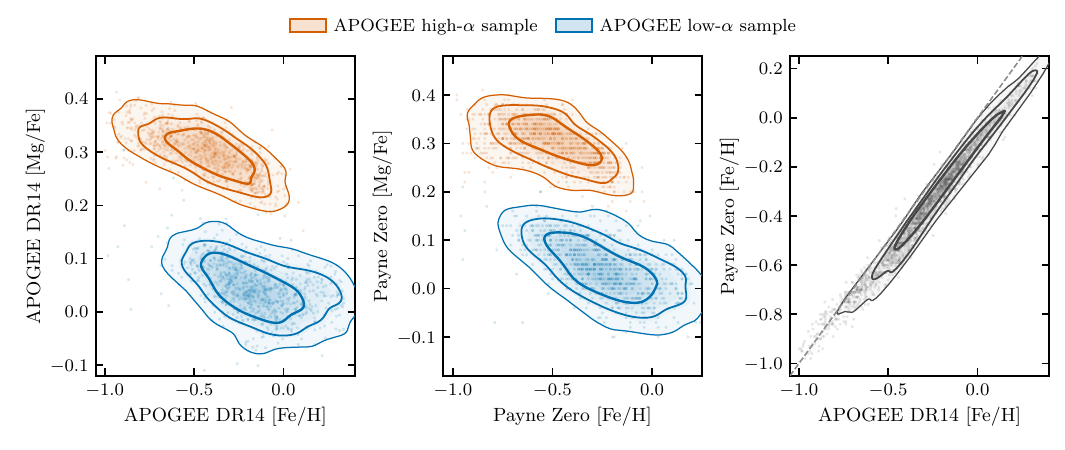}
\caption{Chemical structure of the balanced 1,600-target APOGEE sample. The first two panels show $[\mathrm{Mg}/\mathrm{Fe}]$-- $[\mathrm{Fe}/\mathrm{H}]$ from the calibrated DR14 catalog and the corresponding \paynezero\ fits. Orange and blue mark the APOGEE high- and low-$\alpha$ samples. The third panel compares \paynezero\ and APOGEE DR14 iron abundances. The dashed line denotes equality. Contours enclose 50, 80, and 95 percent of the inverse-selection-weighted density. The direct fits retain the two chemical sequences while allowing an abundance zero-point shift.}
\label{fig:apogee_structure}
\end{figure*}

For this experiment, we use a DR14 calibration-derived LSF \citep{Nidever2015,Holtzman2018}, averaged over six representative fibers. A one-time build folds interpolation from the $R_{\rm grid}=300{,}000$ synthesis grid into compact wavelength-dependent weights that remain on the GPU. The operator maps directly onto the APOGEE pixels without constructing a dense convolution matrix. A fiber-, exposure-, or detector-specific LSF requires only a different prepared GPU asset. Broadening, shifting, LSF convolution, and sampling take about $240~\mu\mathrm{s}$ per total--continuum flux pair on an H100, compared with $1.4$~s for native synthesis. The instrument model therefore adds negligible cost. The fitted width $v_b$ combines rotation, macroscopic broadening, and residual profile width not represented by the mean instrument model. The public fitter also provides a GPU-resident rotational kernel, although this demonstration retains one Gaussian width.

Smooth continuum differences remain in the pseudo-continuum-normalized DR14 spectra \citep{Holtzman2018}. At every nonlinear trial, weighted linear least squares profiles a fourth-order Legendre basis independently on each detector \citep{GolubPereyra2003}. The likelihood uses the published uncertainties and masks \citep{Holtzman2018} with a 0.5 percent model floor. Because the combined spectra are assembled on a common rest-wavelength grid \citep{Nidever2015}, the fit begins at $v_r=0$ and solves only a local residual shift.

Initialized atmospheres and direct synthesis drive the observed-spectrum search. Each additional abundance requires one response spectrum, so this cost grows approximately linearly rather than combinatorially with the label count. The physical solver then converges an atmosphere at the selected mixture, and a fresh GPU spectrum supplies any final local correction. The accepted model always uses a physically converged atmosphere and direct synthesis. For this 12-abundance fit, the H100 search takes about 40~s per star, and atmosphere convergence adds about 30~s on multicore CPUs.

As a proof of concept, we select 1,600 giants, balanced across the high- and low-$\alpha$ APOGEE sequences and metallicity \citep{Nidever2014,Hayden2015}, and reweight the population contours to represent the parent sample. Figure~\ref{fig:apogee_spectrum} first tests the calculation in pixel space. Across all three detectors, the forward model follows the line forest while the profiled continuum absorbs only smooth residual structure. This establishes that the instrument operators and many-element spectrum can be fitted together.

We then test whether the independently fitted abundances preserve population structure. Figure~\ref{fig:apogee_structure} retains the two principal $[\mathrm{Mg}/\mathrm{Fe}]$ sequences over the sampled metallicity range rather than collapsing them into one bulk-metallicity trend. The sample is balanced on the DR14
alpha abundance, with 800 stars from each sequence, so two groups appear in that
figure by construction. Whether an independent measurement keeps them apart is a
separate question. Measured in our fits the two groups differ by 0.22 to 0.24~dex
in $[\mathrm{Mg}/\mathrm{Fe}]$ within metallicity bins, against a
within-sequence scatter of 0.03 to 0.04~dex, so the separation is resolved at six
to seven times that scatter.

\begin{figure*}[t]
\centering
\includegraphics[width=\textwidth]{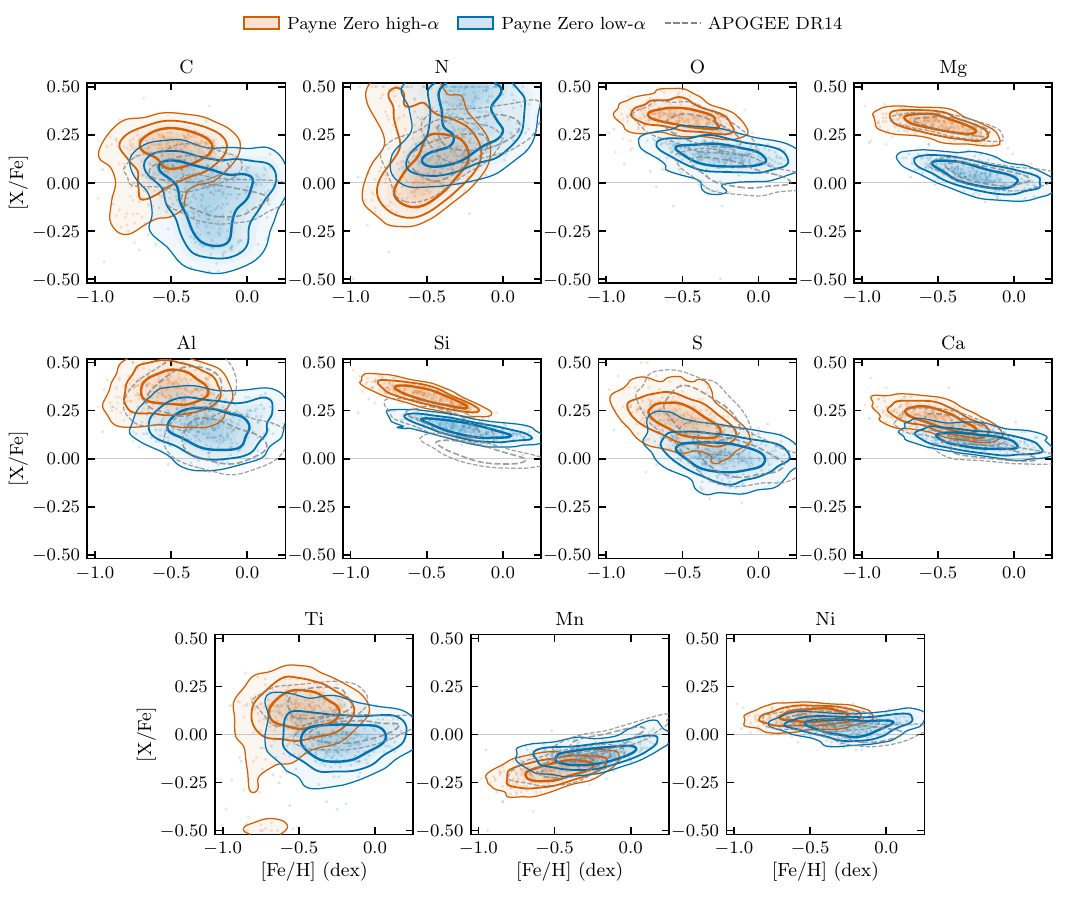}
\caption{Elemental-abundance distributions from the balanced 1,600-target \paynezero\ fit. Each panel shows one $[\mathrm{X}/\mathrm{Fe}]$ ratio against $[\mathrm{Fe}/\mathrm{H}]$. The element is named above the panel. Orange and blue points and contours denote stars drawn from the APOGEE high- and low-$\alpha$ samples, respectively. Colored contours enclose 50, 80, and 95 percent of the inverse-selection-weighted fitted density. Grey dashed contours show the corresponding APOGEE DR14 sample at 50 and 80 percent, computed from the same stars and weights. The fitted abundance planes show coherent element-dependent sequences rather than a single bulk metallicity coordinate.}
\label{fig:apogee_elements}
\end{figure*}

Star by star, the scatter against DR14 ranges from 0.028~dex for nickel to
0.158~dex for nitrogen. Measured the same way on the same stars and separately
for each alpha sequence, the within-sequence scatter of the two analyses has a
median ratio near unity across the twenty-four comparisons. Ours is tighter by
factors of 1.3 to 1.7 for silicon, sulphur, oxygen, and manganese, and matches
DR14 to within 15 percent for iron, magnesium, calcium, and nickel. Carbon,
nitrogen, and titanium are looser by factors of 1.3 to 2.5, and the first two of
those are set by molecular bands.

Figure~\ref{fig:apogee_elements} extends the comparison element by element. O, Si, S, Ca, and Ti retain chemical separation between the two input populations, although their absolute zero points differ.
The nearly rigid Si offset is consistent with a scale difference because DR14 applies empirical temperature and zero-point calibrations to individual abundances, with Si among its larger adjustments, while we apply no APOGEE zero-point correction \citep{Holtzman2018}. Mn rises toward higher metallicity, consistent with metallicity-dependent yields and a substantial Type Ia supernova contribution, while Ni broadly tracks Fe \citep{Weinberg2019}. C and N remain separately constrained, preserving the mass- and evolution-dependent C/N information measured in APOGEE red giants \citep{CharbonnelLagarde2010,Masseron2015,Martig2016}. The final abundance corrections are typically at the hundredth-dex level and leave this population morphology unchanged.

The abundance offsets do not depend on which ASPCAP release provides the reference.
Comparing the same fitted labels against DR17, which uses MARCS atmospheres and
Turbospectrum in place of the DR14 analysis, gives a median absolute difference
of 0.051~dex against 0.055~dex for DR14, while the two releases differ from each
other by 0.024~dex \citep{Abdurrouf2022}.

The learned initializer enters only the fast stage, so the displacement
between the fast solution and the accepted one measures what it contributes to a
reported label. Over the 1,600 accepted stars that displacement has a median of
zero in every coordinate, a half width below 0.015~dex, and a largest single
value of 0.16~dex. Every star converges
on one or two exact atmospheres against a permitted ceiling of eight, and
repeating the fit for twenty of them while varying that ceiling and the
improvement threshold required for each solve holds the median effective
temperature within 6~K.

The line-list transfer is independently measurable. In a matched control sample the calibration lowers the mean chi-square per valid pixel, $Q$, for every star. The control holds eleven giants spanning 4300 to 5200~K in effective temperature, 1.3 to 2.5 in $\log g$, and $-0.95$ to $+0.06$ in $[\mathrm{Fe}/\mathrm{H}]$, with six drawn from the high-$\alpha$ sample and five from the low-$\alpha$ sample. The median $Q$ falls from 3.32 to 1.56, with a median paired reduction of 53 percent.

The calibration was also applied unchanged to 26 stars outside the two
standards it was fitted to, fifteen dwarfs near solar metallicity and eleven giants between
$[\mathrm{Fe}/\mathrm{H}]=-1.6$ and $-2.5$. Each was fitted twice, on the base
catalog and with the calibration, and the mean chi-square per pixel falls for 24
of the 26. The benefit follows how much line-list error the spectrum carries. For the dwarfs it falls from 5.21 to 2.38, close to the matched control, while
for the metal-poor giants it falls from 1.60 to 1.47 in 9 of the 11, since
weaker lines leave less to correct and three already sit below one. The
surface gravity offset against the catalog falls from 0.38 to 0.16~dex, and for
the dwarfs alone from 0.37 to 0.03~dex, while the effective temperatures move a
further 38~K below the catalog scale along the same degeneracy. The most metal-poor stars reach the atmosphere support floor at
$[\mathrm{M}/\mathrm{H}]=-2.5$.

The calibrated line list improves the pixel-level objective without forcing agreement with the ASPCAP abundance scale. The remaining zero-point differences can reflect the atmosphere physics, solar mixture, line selection, masks, and continuum treatment. Together, the spectral fit and the element-by-element sequences show that a direct physical forward model can recover coherent survey chemistry without a label-to-flux spectral emulator. Establishing a calibrated abundance scale would require a broader standard-star set and survey-specific line curation beyond this proof of concept.

The adopted stellar-parameter scale sets part of that zero point.
Appendix~\ref{app:stellar_label_sensitivity} repeats the fit with
$T_{\rm eff}$, $\log g$, and $\xi$ free rather than retained from DR14. The
inverse-selection-weighted medians then move by $-200$~K, $+0.17$~dex, and
$-0.32$~km~s$^{-1}$, and the abundance zero points follow, by up to about
0.2~dex for individual elements while the element sequences persist.

The matched control above separates the line calibration from this shift.
Repeating those fits with the base catalog moves the median temperature
displacement by 3~K, from $-219$~K to $-216$~K, so the temperature offset does not
come from the calibration. The gravity and microturbulence displacements do
respond, by 0.14~dex and 0.40~km~s$^{-1}$, and the abundance zero points move by
up to 0.22~dex.

The fit also returns a formal uncertainty. The weighted response matrix
assembled at the accepted solution gives a Gauss--Newton covariance whose
worst-constrained direction corresponds to 0.04~dex in an abundance ratio, 57~K
in effective temperature, and 0.08~dex in surface gravity, with the
best-constrained combinations an order of magnitude tighter. These follow from
pixel noise alone. The zero-point shifts of up to 0.2~dex above, from the adopted
stellar-parameter scale and from the choice of line list, are several times
larger. The uncertainty that matters for a survey catalog is therefore set by
those choices rather than by the formal value, which is why survey pipelines
calibrate abundance uncertainties against the scatter of cluster members
\citep{Holtzman2018}.

\begin{figure}[t!]
\centering
\includegraphics[width=\columnwidth]{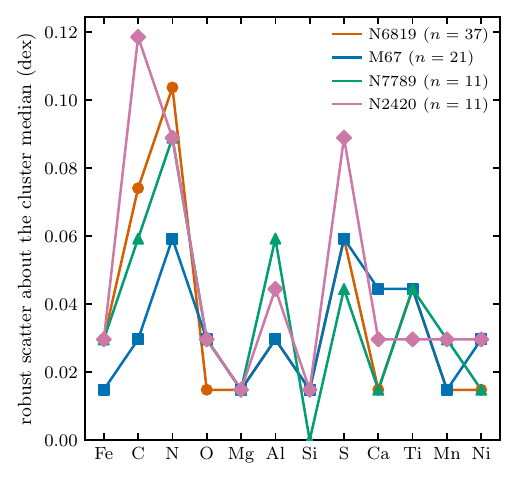}
\caption{Robust scatter of each abundance coordinate about its own cluster median, for the cluster giants fitted with the twelve coordinates used for the survey sample. One line per cluster. No absolute abundance scale enters, since every star is compared with its own cluster.}
\label{fig:m67_homogeneity}
\end{figure}

A cluster tests the recovered chemistry without an absolute scale. Of 105 members in five open clusters, 21 carry the DR14 high-persistence flag and one has no ASPCAP parameters. Both cuts are applied before fitting, so the sample is fixed by the criteria rather than by which fits converge. The remaining 83 span 4100 to 5200~K and were fitted with the same twelve abundance coordinates used for the survey sample, so the recovered chemistry is compared element by element. Across the four clusters with at least five members, which hold 80 of them, the robust scatter about each cluster median is 0.030 dex in iron and, pooled over the clusters, reaches at most 0.089 dex across the eleven $[\mathrm{X}/\mathrm{Fe}]$ coordinates. Nitrogen is the loosest at 0.089 dex, as expected where first dredge-up varies along the giant branch. Figure~\ref{fig:m67_homogeneity} shows both.

\section{Discussion}\label{sec:discussion}

Direct ab initio fitting changes the role of two learned approaches in stellar inference. Spectral emulators interpolate libraries of physical spectra, while data-driven flux models learn from survey observations. Both classes insert a learned label-to-flux relation between the stellar labels and the spectrum being evaluated. When synthesis takes seconds and a converged atmosphere takes seconds to tens of seconds, the physical forward model can instead remain inside inference, line calibration, and validation. Learned atmosphere models can accelerate exploration while physical convergence still determines the accepted atmosphere and spectrum. This changes how we use emulators and how we investigate the observation--model gap.

\subsection{Direct synthesis and the role of spectral emulators}
\label{sec:discussion_emulators}

A physically consistent abundance analysis should vary the relevant stellar labels together and fit the full spectrum. An element changes more than the lines assigned to it. Magnesium changes the electron supply and therefore the H$^-$ continuum. Molecular balance and line blanketing redistribute flux among species and can alter the atmospheric structure \citep{John1988,Gustafsson2008,Meszaros2012,Ting2018Oxygen,Matsuno2024}. Measuring each abundance only from its named features while holding the rest of the model fixed misses these coupled responses.

The difficulty is dimensionality. Each abundance expands the label space, and regular interpolation grids become rapidly sparse \citep{Ting2016}. Probabilistic and neural emulators such as Starfish, The Payne, and later neural architectures replaced that grid with smooth interpolation of physical spectra \citep{Czekala2015,Ting2019,Rozanski2025a}. We introduced The Payne to make simultaneous full-spectrum inference from self-consistent ab initio models computationally practical. A neural emulator was the means of accelerating that calculation, not its scientific objective. The name nevertheless became shorthand for the emulator rather than the physical fitting framework.

\paynezero\ returns to that original goal without a learned label-to-flux interpolator. ``Zero'' denotes zero label-to-flux emulation layers rather than the absence of every learned component. Every trial spectrum is calculated by direct synthesis for the requested labels and mixture, and the accepted solution is checked with a physically converged atmosphere. One learned component does enter the search. The initializer of Section~\ref{sec:atmosphere_emulator} predicts a starting atmosphere from the stellar labels, so during the fast stage the depth structure behind each trial spectrum comes from a network rather than from a converged solve. That component is a label-to-atmosphere model and not a label-to-flux model. It is never accepted as an atmosphere, it is never fitted to an observed spectrum, and every reported label comes from a physically converged atmosphere with a direct synthesis at those labels. 

An error in a flux emulator enters the likelihood directly and is absorbed into the fitted labels. An error in the initializer costs additional physical passes or a failure to converge, and the convergence test still decides what is accepted. Full-spectrum fitting can therefore retain the coupled response of all labels without constructing a high-dimensional spectral training grid. The APOGEE demonstration frees 12 of the 81 abundance coordinates the forward model accepts, following the information the H band carries.

The demonstration is a point-estimate fit rather than a sampled posterior.
A sampler needing $10^4$ to $10^6$ likelihood evaluations would take hours to
hundreds of hours per star at the synthesis times reported here, so a
label-to-flux emulator retains a large advantage for that mode of inference. The
advantage is reinforced because the atmosphere-to-spectrum path is not yet a
single differentiable graph and so cannot supply gradients to a Hamiltonian
sampler. 

Sampling is not the only route to an uncertainty. At the
signal-to-noise of a survey spectrum, the curvature of the objective at the
optimum already provides a useful Gaussian estimate, and the weighted response
matrix assembled for the Gauss--Newton step supplies that curvature without any
further synthesis. 
Survey pipelines go further and calibrate their
abundance uncertainties empirically, from the scatter among cluster members
taken to be chemically homogeneous \citep{Holtzman2018}. That route is equally
open to a direct fit and needs no sampled posterior. Direct fitting covers the survey applications demonstrated
here. Sampling remains the setting in which an emulator is the practical choice,
in particular at low signal-to-noise, near a parameter bound, or where the
posterior departs appreciably from a Gaussian.

\subsection{Scope of the demonstration}
\label{sec:discussion_scope}

The line-list calibration of
Section~\ref{sec:linecal} is a separate matter, since it is derived over
1500--1700~nm and does not carry to optical transitions. The calibration
machinery applies unchanged, but an optical analysis would require its own
solution. We defer that application to a subsequent paper.

The demonstrated regime is one-dimensional LTE Kurucz physics. The speedup
in Section~\ref{sec:synthesis} rests on wavelength independence at a fixed
atmosphere, and non-LTE statistical equilibrium removes exactly that property,
because level populations couple to the radiation field at every frequency that
connects to a transition. A non-LTE calculation would therefore not reduce to a
single wavelength batch. Its natural parallel unit is instead the set of depths,
levels, and transitions evaluated inside an outer iteration on the populations,
so the wavelength-parallel kernel developed here becomes the inner operation of
that loop rather than the whole calculation. 
We expect the same reorganization to
remain useful in that setting, since the inner opacity and transfer work is still
large and independent, but the gain would be smaller than the factors reported
here and we have not demonstrated it. Departures from LTE also change the
emergent line strengths themselves, so a non-LTE forward model would alter the
recovered abundances and the line calibration rather than the runtime alone.

\subsection{From learned flux relations to physical calibration}
\label{sec:discussion_calibration}

Data-driven label-to-flux relations address the observation--model gap from the other direction. The Cannon, for example, learns directly from spectra and reference labels rather than evaluating an explicit physical forward model. This empirical route can reproduce repeatable structure missing from a synthetic library. Its caveat is attribution. Correlated labels in the training sample can produce a useful spectral predictor without necessarily identifying the physical process responsible for a feature.

When the goal is to improve the forward model, direct ab initio fitting tests the proposed physical cause. Structured residuals can arise from line data, missing opacity, and broader synthetic--observed mismatch \citep{Czekala2015,OBriain2021}. Atmosphere, continuum, telluric, and reduction errors can produce related residuals, but they enter the calculation in different places. A differentiable physical model exposes these inputs and tests a proposed cause through opacity and radiative transfer.

The line-list calibration demonstrates this route. We optimize more than $10^5$ coupled oscillator-strength and damping corrections at once, including blended transitions, and obtain the shared Sun--Arcturus solution in about one minute on one H100. These values are effective standard-star calibrations rather than laboratory measurements. Their performance on a frozen APOGEE control set tests transfer without using ASPCAP labels to determine the corrections.

The correction is a flexible model by construction, with more than $10^5$
free values fitted to two stars. What separates it from a label-to-flux relation
is where the parameters live. A correction to an oscillator strength or a damping
constant belongs to the transition rather than to the star, so one value acts on
every spectrum in which that line appears and is constrained by every pixel the
line touches. It also has a laboratory counterpart that can be measured without
any spectrum. Section~\ref{sec:linecal} reports how little of that freedom the
fitted solution uses and how the corrections behave on an independent
observation.

The calibration is degenerate with the abundances it is later used to measure. Corrections are grouped within a species and ion stage while the standard-star abundances stay fixed, so an error in the abundance of a species is absorbed almost exactly into the oscillator strengths of its lines. Only saturated lines separate the two, and two standards supply little of that leverage. The same holds along the microturbulence, damping, and continuum chain.

The matched control of Section~\ref{sec:apogee} bounds the size of this
effect. Fitting the same eleven giants with the base catalog and with the shared
calibration moves the abundance coordinates by a median of 0.12~dex and by up to
0.22~dex. The released corrections are therefore an effective correction tied to
the adopted solar and Arcturus abundance scale and to the atmosphere, continuum,
and microturbulence treatment used here. Because the fit absorbed those choices, the
corrections are not a stand-alone atomic data set. An analysis that adopts a
different solar mixture or a different microturbulence prescription breaks the
cancellation the fit relied on, and the residual error reappears in its
abundances. The element-dependent zero points in
Figure~\ref{fig:apogee_elements} are consistent with this degeneracy acting
alongside the DR14 empirical calibrations we do not apply.

A production calibration could use several benchmark standards \citep{Heiter2015,Jofre2015,Smith2021}. We propose solving their shared line parameters while allowing each star's labels, continuum, and instrument response to vary. The appropriate balance among standards is not determined by the optimizer. The Sun and Arcturus probe different pressure and molecular regimes, so the equal weighting used here demonstrates capacity rather than defining a universal calibrated catalog. A production analysis should choose its standards, weights, laboratory priors, and held-out tests for the stellar population being analyzed.
Large opacity revisions would then trigger a new atmosphere calculation and another line-data update. This outer iteration preserves physical coupling and reveals discrepancies that line calibration cannot repair, including failures of the atmosphere iteration itself.

\subsection{Atmosphere boundaries and reduced solvers}
\label{sec:discussion_fixed_point}

The parity tests span hot-dwarf, solar, red-giant, and K-dwarf regimes, while the survey fit is restricted to giants. These results do not imply that one numerical strategy will remain efficient across every stellar regime. Two edge cases identify where a different representation or solver may be needed.

The first is the sampling dimension of the atmosphere initializer. The direct-abundance model compresses six correlated depth profiles into a compact PCA representation, but its input still spans three stellar parameters together with 81 independently varied abundances. The resulting 84-coordinate input is sparse relative to its dimension. The same sampling problem limits high-dimensional model grids. Correlated atmospheric responses nevertheless make the initializer useful. Future initializers could use coordinates tied more closely to electron donation, molecular balance, and line blanketing, or learn such groupings with a chemistry-aware encoder. The initializer need only propose a starting state because the physical solver still establishes the accepted atmosphere at the requested individual abundances.

The second is a change of numerical regime in cold molecular atmospheres. Some cooler high-gravity controls develop growing flux and heating residuals even from a nearby initial profile, while other nearby starts converge. Convergence therefore becomes more sensitive to the initial structure across this regime. Both the original Kurucz calculation and \paynezero\ fail the same convergence criterion in some of these controls, which points to the physical iteration rather than an emulator-specific error.

Small temperature changes in this regime reorganize molecular equilibrium \citep{BarklemCollet2016}, opacity, the equation of state, and the adiabatic gradient together. Convection can also reach the layers that set the emergent flux \citep{Gustafsson2008}. One iteration can then shift the Rosseland opacity, column-mass mapping, and convective boundary enough that the next correction grows rather than shrinks. The usual fixed-point iteration has lost local stability.

For a reduced plane-parallel formulation, we choose temperature $T(\tau_{\rm R})$ and monotonic column mass $m(\tau_{\rm R})$ as the two iterated profiles. At fixed $T_{\rm eff}$, $\log g$, $\xi$, abundances, and Rosseland grid, hydrostatic balance and the equation of state then recover pressure, density, electron density, and populations \citep{HubenyMihalas2014}. Opacity sets the relation between column mass and Rosseland depth, while radiative transfer sets the acceleration. A reduced solver would enforce two coupled root conditions, energy balance and consistency between column mass and Rosseland optical depth,
\begin{equation}
\begin{aligned}
\mathcal{R}_{T}[T,m]
&=F_{\rm rad}[T,m]+F_{\rm conv}[T,m]
-\sigma_{\rm SB}T_{\rm eff}^{4}=0,\\
\mathcal{R}_{m}[T,m]
&=\frac{\mathrm{d}\tau_{\rm R}}{\mathrm{d}m}-\kappa_{\rm R}[T,m]=0 .
\end{aligned}
\end{equation}
\noindent These residuals motivate a solver in $T$ and $m$ rather than an initializer for all six stored profiles. Recomputing the other four fields did not restore convergence from learned cool-dwarf $T$ and $m$ profiles, so the instability is not simply caused by predicting too many fields.

A native two-profile solver must recompute the equation of state, opacity, transfer, and convection for each trial $T$ and $m$, then solve both residuals together. In exploratory calculations, nearby physical models estimated how both residuals change with $T$ and $m$ and supplied local corrections. The residuals fell, but the calculation required more complete atmosphere evaluations than the learned six-field start followed by the physical solver iteration. We therefore retain the latter as the default. A two-profile solver remains a possible fallback if its response can be estimated without many full physical evaluations. Different stellar regimes may ultimately require different validated correctors. Choosing when to switch regimes, while preserving the same physical acceptance tests, becomes a workflow problem. It motivates software boundaries that expose physical state, convergence, and validation rather than hiding them inside one opaque calculation.

\subsection{Readable software as a scientific interface}

Switching safely between numerical regimes requires the physical state, residuals, and acceptance tests to remain visible. The original Kurucz workflow encodes extensive physics in compact Fortran programs connected by staged files \citep{Kurucz2005}. During reimplementation, we found that terse identifiers, long routines, and implicit shared state made individual conventions difficult to isolate. Assumptions about line placement, opacity accumulation, or state transfer can therefore be difficult to audit and extend even when they change the final spectrum.

\paynezero\ replaces terse identifiers and implicit handoffs with named physical arrays, explicit units, and stage-local interfaces. These choices expose numerical conventions for audit, reproduction, teaching, and extension.

The same explicit interfaces make \paynezero\ suitable for future agentic workflows. Model Context Protocol tools or task-specific skills could expose atmosphere solving, spectrum synthesis, line calibration, and validation as typed operations with recorded inputs, outputs, and provenance.\footnote{See the Model Context Protocol specification at \url{https://modelcontextprotocol.io/specification/2025-06-18/server/index}.} An agent could inspect convergence and flux residuals, recognize that a cool high-gravity model has entered a different numerical regime, select a prevalidated solver or parameterization, and launch the relevant diagnostic tests. The agent would orchestrate these routes rather than certify its own result. Deterministic physical, numerical, and provenance gates would still decide acceptance.

\section{Conclusion}\label{sec:conclusion}

Spectral emulators and data-driven flux models addressed two longstanding problems in stellar spectroscopy. Physical calculations were too slow to repeat inside inference, and their spectra did not reproduce every observed feature. The resulting label-to-flux relations made survey analysis practical, but spectral emulators add interpolation error, while data-driven models can inherit correlations among their training labels. We show that this intermediate layer is no longer required for point-estimate fitting in the demonstrated one-dimensional LTE regime, while sampled inference at $10^4$ to $10^6$ likelihood evaluations remains a setting where an emulator is the practical choice.

\paynezero\ changes this tradeoff by reorganizing the calculation for modern processors without changing its physical target. GPU-native synthesis reduces a 300--1000~nm solar calculation at $R_{\rm grid}=300{,}000$ from about 760~s in the original Kurucz programs to 14~s on one H100, while the APOGEE interval takes about 1~s. Each physical atmosphere iteration gains a factor of 1.4--2.0 from algorithmic reorganization and a further 3.5--4.5 from sixteen CPU threads, six to seven combined. Only the algorithmic term applies to bulk throughput, where either code can distribute stars over cores. A learned starting atmosphere reduces the required iteration count by about another factor of three in the controlled tests. The resulting spectra remain in practical parity with the original calculations. The atmosphere and synthesis calculation can therefore remain inside inference, where their assumptions and residuals are evaluated directly.

The same computational structure turns the observation--model gap into a tractable physical optimization problem. Automatic differentiation updates more than $10^5$ coupled oscillator strengths and damping parameters together. The shared Sun--Arcturus calibration takes about one minute on one H100, replacing a transition-by-transition task with one joint radiative-transfer calculation.

These capabilities meet in our APOGEE demonstration. GPU-resident velocity shifting, broadening, wavelength-dependent LSF convolution, and detector sampling add negligible cost to synthesis. For APOGEE giants, the fit solves 12 abundance coordinates together with the instrument and continuum terms directly from reduced one-dimensional spectra. The H100 search takes about 40~s per star, while physical atmosphere convergence runs independently on multicore CPUs. The recovered chemistry shows coherent element-dependent trends across all eleven fitted abundance ratios for stars spanning the giant branch and red clump. That demonstration is confined to the H band, while most of the surveys motivating this work are optical, where the residuals in Figure~\ref{fig:synthesis_overview} grow toward the blue but remain at or below the $10^{-3}$ level longward of 400~nm.

The resulting computational scale changes what is practical for a survey. The measured direct-synthesis search for $10^5$ spectra corresponds to about one thousand GPU-hours and a few thousand US dollars at typical academic cloud rates. Independent stars can be distributed without changing the calculation. Atmosphere convergence uses multicore CPU workers and can run independently of the GPU synthesis tasks. GPU demand is therefore set mainly by the direct-synthesis search and direct spectra rather than by the atmosphere iteration.

Stellar spectroscopy has moved from individually operated physical programs, through precomputed grids and learned approximations, to surveys of millions of stars. Modern hardware now allows the physical calculation to return to the inference loop at that scale. This does not remove departures associated with three-dimensional convection or non-LTE line formation \citep{Asplund2005,Nordlund2009,Magic2013,BergemannNordlander2014,Amarsi2019}, nor limitations in line data or instrument modeling. It makes them visible targets for improvement. Atmospheres can be recomputed at the requested mixture, spectra can be synthesized inside the optimizer, and atomic data can be calibrated through the same radiative-transfer calculation. Physical forward models can again be the working foundation of stellar inference rather than only the training source for an intermediate learned flux model.

\section*{Acknowledgments}

\begin{figure*}[t!]
\centering
\includegraphics[width=\textwidth]{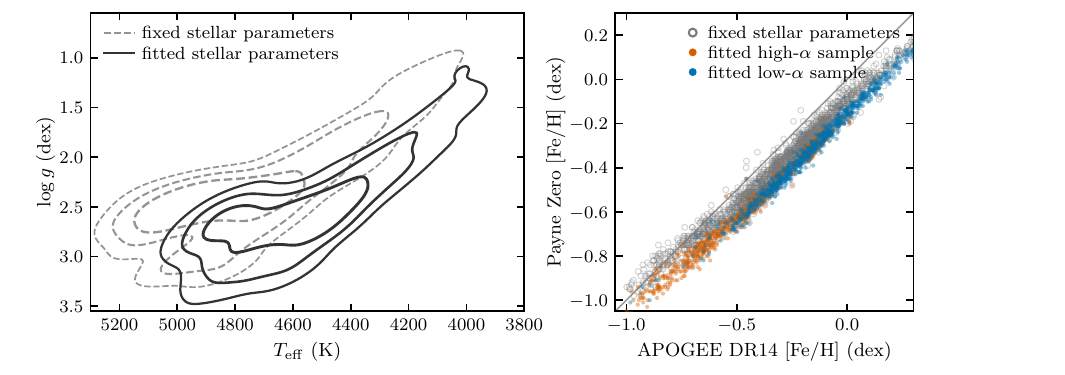}
\caption{Sensitivity to fitting $T_{\rm eff}$, $\log g$, and $\xi$ together with the 12 abundance coordinates. Left, inverse-selection-weighted contours for the combined sample. Gray dashed curves retain the DR14 stellar parameters and dark solid curves show the final physically converged fit. Right, fixed-parameter iron abundances are gray open circles and the final values are colored by the APOGEE high- and low-$\alpha$ samples.}
\label{fig:appendix_stellar_label_sensitivity}
\end{figure*}

For YST, this project was also a hobby and an informal test of the Pareto frontier of modern large-language-model agents for scientific software. Attempts beginning in 2024 could translate isolated files but could not preserve the logic of a large coupled codebase. A practical turning point came in December 2025, when Claude Opus 4.5 made careful translation of Kurucz routines into Python tractable. EK carried out much of the initial painstaking translation into PyKurucz and developed its early Numba implementation, leading to the pyKurucz study of Kim \& Ting (2026). YST led the subsequent GPU-native redesign, optimization experiments, and Payne Zero development. Progress beyond a brute-force rewrite depended on YST's experience with the Kurucz codes to identify the resident data, kernel boundaries, line selection, and parallel structure. Later development used Claude Opus 4.8 and sustained GPT-5.5 Codex agents, with independent criticism from Gemini 3.1 Pro and GLM 5.2. A rewrite at this scale would not have been practical on the same timescale without these tools. The unsuccessful experiments, including the alternate two-profile atmosphere solver, were equally useful reminders that current agents still require close domain guidance. Artificial general intelligence (AGI) would have made this hobby considerably shorter.

\begin{figure*}[tp]
\centering
\includegraphics[width=\textwidth]{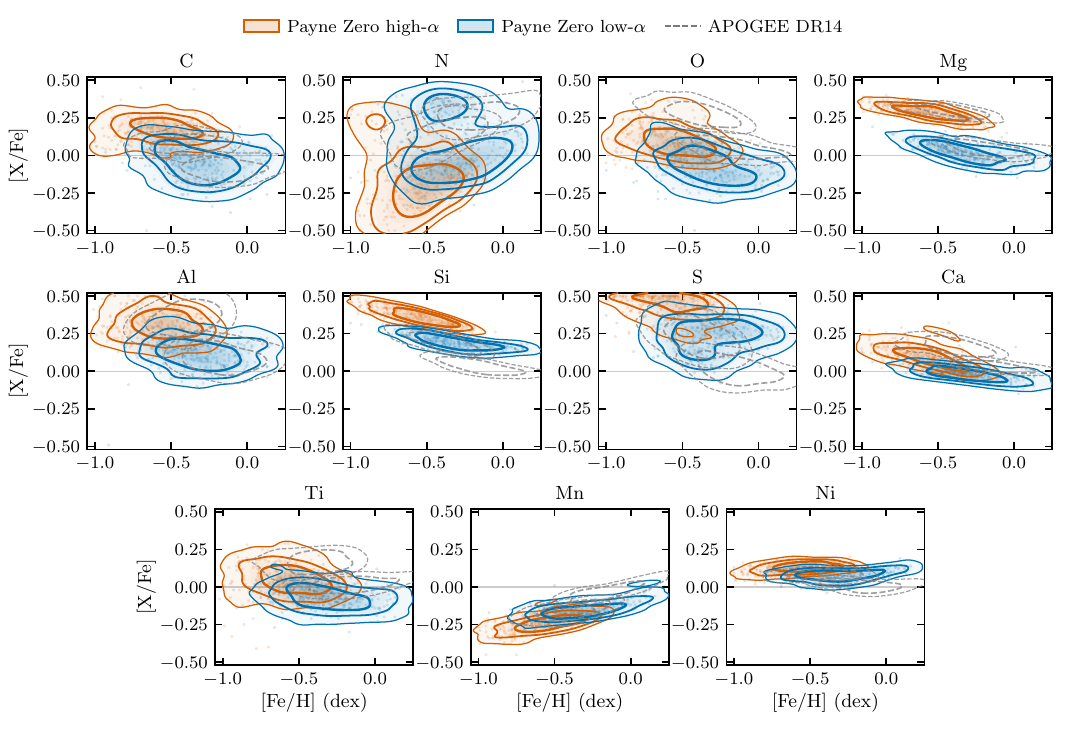}
\caption{Elemental-abundance distributions from the final physically converged fit of $T_{\rm eff}$, $\log g$, $\xi$, and the abundances. The layout and scales match Figure~\ref{fig:apogee_elements}. Colored points and contours show the final values for the two APOGEE samples. Gray dashed contours show the corresponding 50 and 80 percent density levels from APOGEE DR14.}
\label{fig:appendix_abundance_sensitivity}
\end{figure*}

\begin{figure*}[tp]
\centering
\includegraphics[width=\textwidth]{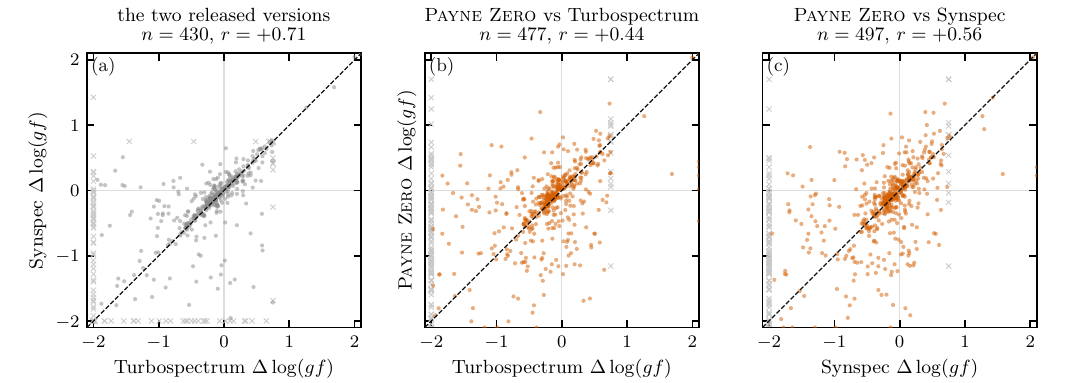}
\caption{Astrophysical oscillator-strength corrections for transitions
shared with the two versions of the APOGEE line list released in Data Release 17
\citep{Smith2021,Abdurrouf2022}. Both were tuned against the Sun and Arcturus
from the same starting atomic data and differ in the synthesis used, so panel
(a) measures how much such a calibration depends on the synthesis alone. Panels
(b) and (c) place the corrections of this work against each version in turn.
Dashed lines mark equality. Each panel shows only transitions that both of its
analyses adjusted. Grey crosses are corrections sitting exactly on a bound of
the APOGEE calibration, which are clipped rather than fitted and are excluded
from the quoted statistics. The bounds at $-2$ and $+0.75$~dex apply to
transitions with no known oscillator-strength uncertainty \citep{Smith2021}.}
\label{fig:linelist_comparison}
\end{figure*}

The agents contributed code translation, refactoring, tests, diagnostics, research-log maintenance, and prose revision. The scientific framing, Introduction, Discussion, and Conclusion began from human drafts. Some technical passages began from agent summaries of the code and research logs. The authors reviewed the retained code, calculations, figures, citations, and text and take responsibility for the results. Unexpected personal circumstances also gave YST the uninterrupted time needed to finish this project.

YST acknowledges support from a Humboldt Research Fellowship from the Alexander von Humboldt Foundation.

\begin{appendix}

\section{Sensitivity to Fitting the Stellar Parameters}
\label{app:stellar_label_sensitivity}

The main APOGEE experiment retains the photometrically calibrated $T_{\rm eff}$ and asteroseismically calibrated giant $\log g$ published in DR14, and fixes $\xi$ to isolate the chemical and instrumental fit \citep{Holtzman2018}. We test that choice with the same official spectra, sample, line calibration, instrument operator, masks, uncertainties, and continuum treatment. Starting from the main abundance solution, we fit those three stellar parameters together with the same 12 abundance coordinates and converge the final physical atmosphere. Figures \ref{fig:appendix_stellar_label_sensitivity} and \ref{fig:appendix_abundance_sensitivity} compare the resulting stellar and chemical distributions with the fixed-parameter solution and APOGEE DR14.

The full physical refit gives inverse-selection-weighted median changes of $-200$~K in $T_{\rm eff}$, $+0.17$~dex in $\log g$, $-0.32$~km~s$^{-1}$ in $\xi$, and $-0.06$~dex in $[\mathrm{Fe}/\mathrm{H}]$. Among the 694 stars in the initial red-clump concentration, the median changes are $-220$~K and $+0.17$~dex. The giant-branch ridge and red-clump concentration remain visible, but their absolute location moves with the adopted stellar-parameter scale.

The median abundance changes range from $-0.23$~dex for nitrogen and $-0.22$~dex for oxygen to $+0.22$~dex for sulfur. Magnesium changes by $-0.03$~dex. The abundance sequences persist while their zero points respond to the stellar parameters. The cooler spectroscopic temperature scale therefore propagates into element-dependent abundance zero points.
A joint fit to the spectrum, broadband spectral energy distribution, and parallax can anchor $T_{\rm eff}$ and $\log g$ and absorb this scale difference in a production analysis \citep{Cargile2020}. We do not include that external information in this proof of concept.

\section{Comparison with a Published Astrophysical Line List}
\label{app:linelist_comparison}

Section~\ref{sec:linecal} sets our corrections against the two versions of
the APOGEE line list released in Data Release 17. The matching requires the same
species and ion stage and a wavelength agreement within 0.002~nm, and retains only transitions whose
starting oscillator strength agrees with ours to better than 0.01~dex.
The comparisons below use only transitions that both analyses being
compared adjusted by more than 0.01~dex, so lines left at their catalog values
do not enter the correlations. Of the transitions each released version adjusts,
147 of the Turbospectrum and 114 of the Synspec corrections sit exactly on a
bound and are clipped rather than fitted, and we drop those. None of ours lands
on a bound. The correlations below therefore rest on 477 transitions against the
Turbospectrum version and 497 against the Synspec version.

Figure~\ref{fig:linelist_comparison} sets the three sets of corrections
against each other. Because the two released versions differ only in the
synthesis used to tune them, the spread between them measures how much an astrophysical calibration
depends on the synthesis alone. They correlate at $+0.71$ on 430 transitions
with a median absolute difference of 0.07~dex. Astrophysical oscillator
strengths therefore carry a component that depends on the synthesis rather than
on the transition.

We adjust 2,062 of the shared transitions, against 633 in the Turbospectrum
version and 622 in the Synspec version. Our corrections correlate with the
Turbospectrum version at $+0.44$ on 477 transitions and with the Synspec version
at $+0.56$ on 497, with median absolute differences of 0.18 and 0.17~dex. The
typical adjustment is 0.26~dex in this work and 0.27~dex in the Turbospectrum
version.

The analyses differ in several respects. The APOGEE calibration uses MARCS
atmospheres of 56 depth points with Turbospectrum, selects 1,664 atomic lines by
their depth in the standard stars, and holds each oscillator-strength correction
within twice its published uncertainty, or within $-2.0$ and $+0.75$~dex where no
uncertainty is known. Damping is adjusted for 250 strong solar lines and
restricted to 0.4~dex \citep{Smith2021}. We use the Kurucz atmosphere and
synthesis, admit every neutral and singly ionized transition in the interval, and
adopt wider bounds. Admitting every transition rather than a depth-selected
subset is the reason we adjust more of them, and the wider bounds are the reason
our adjustments are not clipped.

\end{appendix}

\newpage

\bibliographystyle{aasjournal}
\bibliography{references}

\end{document}